\documentclass{article}

\usepackage[final,nonatbib]{neurips_2024}
\usepackage[title]{appendix}

\usepackage[utf8]{inputenc} 
\usepackage[T1]{fontenc}    
\usepackage{hyperref}       
\usepackage{url}            
\usepackage{booktabs}       
\usepackage{amsfonts}       
\usepackage{nicefrac}       
\usepackage{microtype}      
\usepackage{xcolor}         
\usepackage{graphicx}
\usepackage{amsmath}
\usepackage[numbers]{natbib}
\usepackage{adjustbox}
\usepackage[bottom]{footmisc}
\usepackage{booktabs}
\usepackage{float}
\usepackage{textcomp}
\usepackage{algorithm}
\usepackage{algpseudocode}
\usepackage{subfig}
\usepackage{bm}
\usepackage{amssymb}
\usepackage{color}
\usepackage{tabularx}
\usepackage{caption}
\usepackage{anyfontsize}
\newcommand{\ra}[1]{\renewcommand{\arraystretch}{#1}}

\newcommand*{\img}[1]{%
    \raisebox{-0.1\baselineskip}{%
        \includegraphics[
        height=2mm,
        width=18mm,
        ]{#1}%
    }%
}
\newcommand\like[1]{\begin{picture}(1,1)
\ifnum0=#1\put(.5,.35){\circle{1}}\else
\ifnum10=#1\put(.5,.35){\circle*{1}}\else
\put(.5,.35){\circle{1}}\put(.5,.35){\circle*{.#1}}
\fi\fi\end{picture}}

\title{SCOREQ: Speech Quality Assessment with Contrastive Regression}

\author{%
  Alessandro Ragano \\
  School of Computer Science\\
  University College Dublin\\
  Dublin, Ireland \\
  \texttt{alessandro.ragano@ucd.ie} \\
  \And
  Jan Skoglund \\
  Google LLC \\
  San Francisco, CA \\
  \texttt{jks@google.com} \\
  \AND
  Andrew Hines \\
  School of Computer Science \\
  University College Dublin \\
  Dublin, Ireland \\
  \texttt{andrew.hines@ucd.ie} \\
}

\begin{document}

\maketitle

\begin{abstract}
In this paper, we present SCOREQ, a novel approach for speech quality prediction. SCOREQ is a \textit{triplet loss function for contrastive regression} that addresses the domain generalisation shortcoming exhibited by state of the art no-reference speech quality metrics. In the paper we: (i) illustrate the problem of L2 loss training failing at capturing the continuous nature of the mean opinion score (MOS) labels; (ii) demonstrate the lack of generalisation through a benchmarking evaluation across several speech domains; (iii) outline our approach and explore the impact of the architectural design decisions through incremental evaluation; (iv) evaluate the final model against state of the art models for a wide variety of data and domains.  The results show that the lack of generalisation observed in state of the art speech quality metrics is addressed by SCOREQ. We conclude that using a triplet loss function for contrastive regression improves generalisation for speech quality prediction models but also has potential utility across a wide range of applications using regression-based predictive models.
\end{abstract}

\section{Introduction}
No-reference speech quality assessment has seen significant advancements in recent years, thanks to supervised learning~\cite{avila2019non,Quality-Net,serra2021sesqa,ragano2021more,mittag21_interspeech,catellier2020wawenets,reddy2021dnsmos,becerra22_interspeech}. Supervised speech quality models learn to map input features (waveform domain, mel spectrograms) to Mean Opinion Score (MOS), a continuous target value derived by averaging individual listener ratings on a predefined Absolute Category Rating (ACR) scale (1=Bad, 2=Poor, 3=Fair, 4=Good, 5=Excellent).

While no-reference deep learning MOS predictors outperform traditional full-reference metrics (e.g, PESQ~\cite{rix2001perceptual}, POLQA~\cite{beerends2013perceptual}, ViSQOL~\cite{chinen2020visqol}, CDPAM~\cite{manocha2021cdpam}), they struggle to generalise to unseen audio degradations. Addressing the domain mismatch in speech quality metrics is urgent given the fast growing research in generative speech e.g., neural speech coding~\cite{zeghidour2021soundstream}, speech enhancement~\cite{lu2022conditional} and speech synthesis~\cite{popov2021grad}. In Figure \ref{fig:domain_mismatch} we show the significant performance gap of no-reference speech quality metrics between in-domain and out-of-domain test sets.

The difficulty of no-reference speech quality models is the attempt to map high-dimensional data $\mathcal{R}^D$ to a monodimensional continuous space. Speech data include several entangled factors that are irrelevant to quality and they make it difficult for models to learn a low-dimensional space e.g., speaker identity, pitch, spoken content, phonetic information, degradation separability, etc. State-of-the-art methods that attempt to improve generalisation are based on either expanding the dataset, developing better architectures, or finetuning pre-trained self-supervised learning (SSL) methods. However, as we will show in this paper, domain mismatch is still a problem when tested on unseen degradations.

\begin{figure}[!t]%
    \centering
    \subfloat[\centering L2 loss \newline \newline NMI($\textbf{x},\blacklozenge,\bullet,\blacksquare,\blacktriangleright$)=0.39 \hspace{-16mm} \newline PC(\img{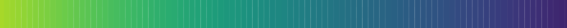})=0.53]{{\includegraphics[width=0.40\textwidth]{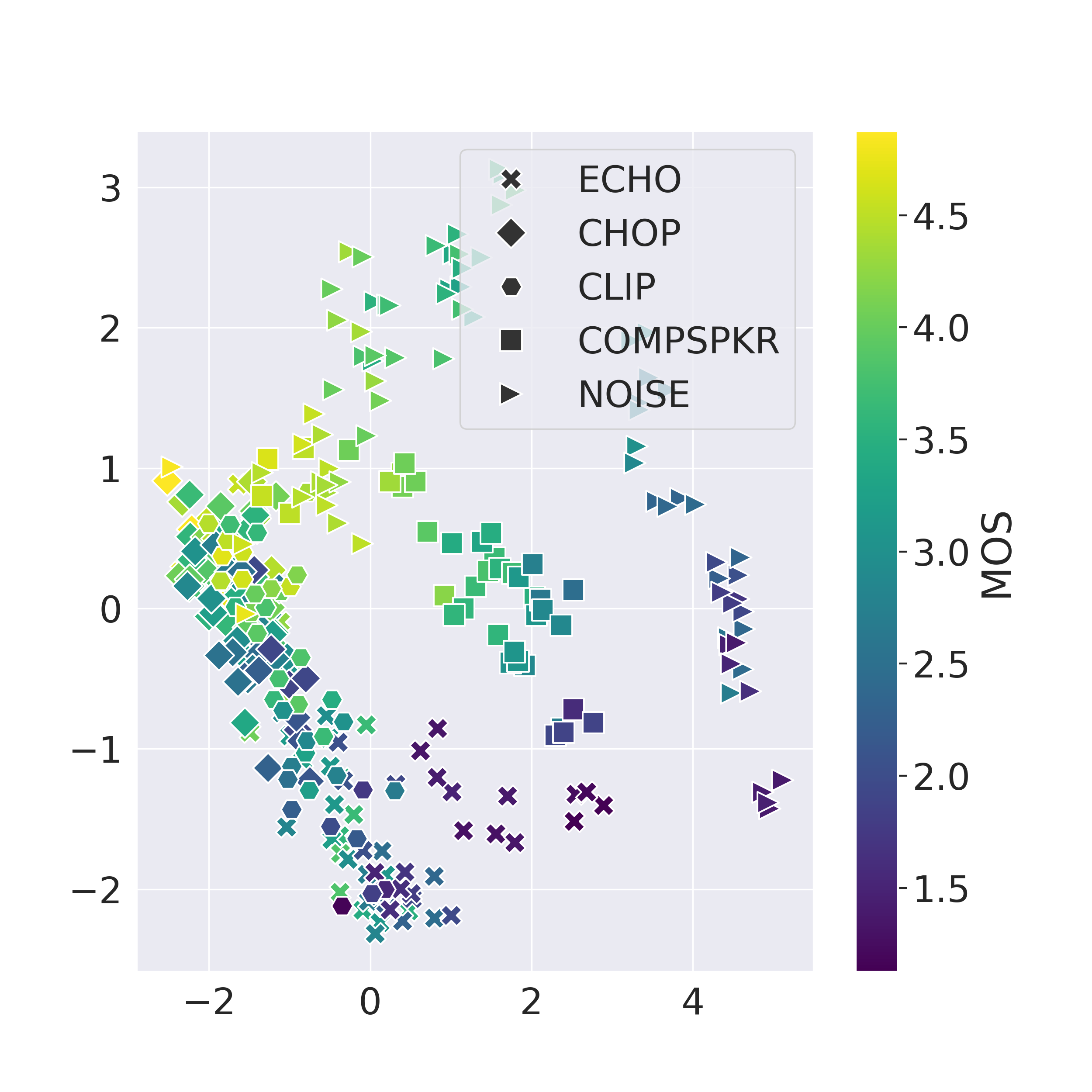} }}
    \qquad
    \subfloat[\centering SCOREQ \textit{(Ours)} \newline \newline  NMI($\textbf{x},\blacklozenge,\bullet,\blacksquare,\blacktriangleright$)=0.11 \hspace{-16mm} \newline  PC(\img{colorbar.png})=0.79]{{\includegraphics[width=0.40\textwidth]{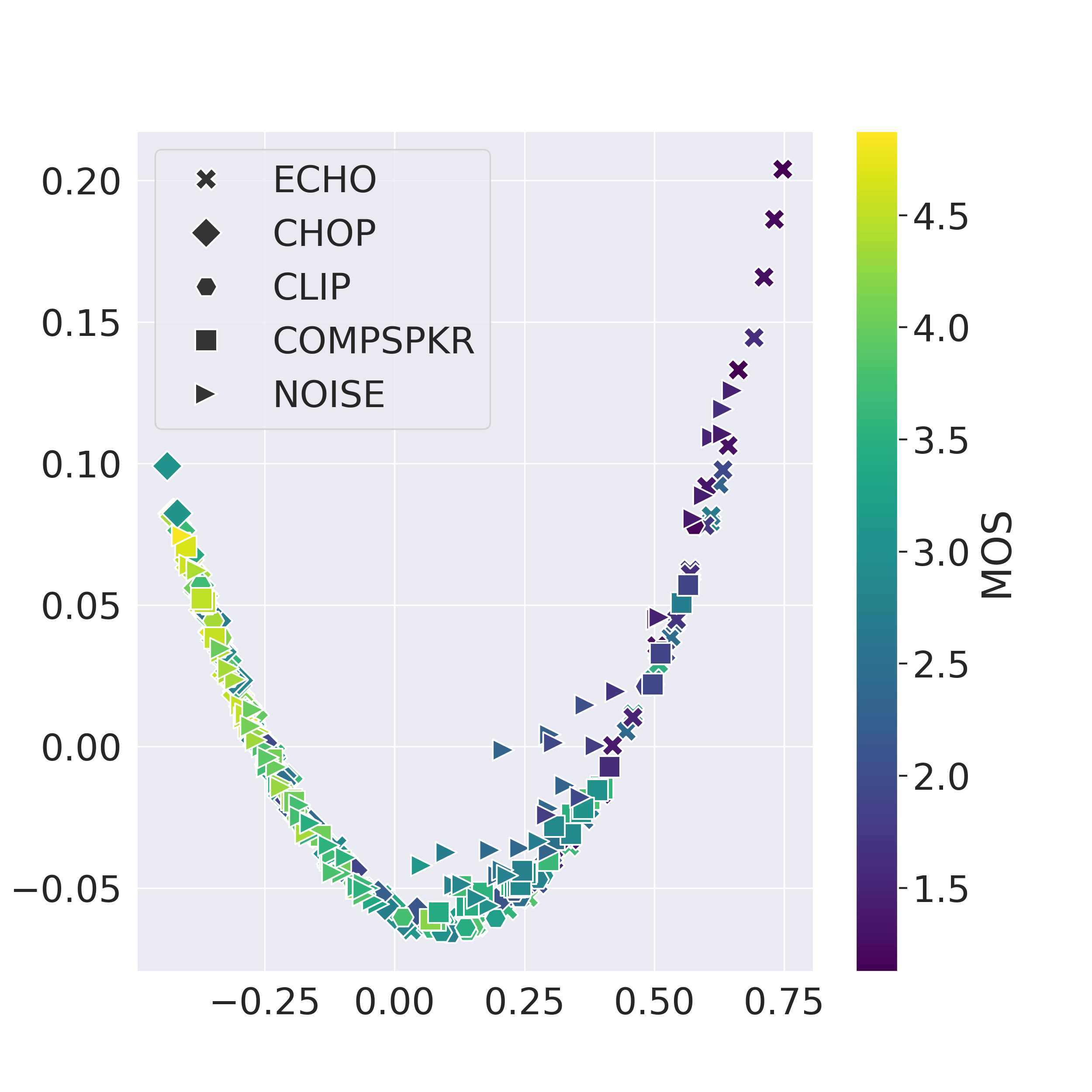} }}
    \caption{Embeddings of L2 loss (a) vs SCOREQ (b) on TCD VOIP data~\cite{harte2015tcd}. Color shows quality labels (MOS) while markers identify the degradations. We compute the Normalised Mutual Information (NMI) between k-Means clusters and degradation labels, as well as the Pearson's Correlation (PC) between embedding distance with respect to random clean speech and MOS targets. Higher NMI indicates representations are clustered based on degradations while higher PC means representations are ordered with respect to MOS targets. Results indicate that the L2 loss embeddings tend to capture degradation information (NMI=0.39, PC=0.53) while SCOREQ quality (NMI=0.11, PC=0.80). See Appendix \ref{appendix:figure1} for more details. }
    \label{fig:embeddings}
\end{figure}

We observe that the majority of no-reference metrics are either trained or finetuned end-to-end by minimising the L2 loss between ground truth MOS and predictions. This approach is simple but it does not induce an ordered representation with respect to the regression targets in the feature space. The same would apply to L1 loss-based regression. Figure \ref{fig:embeddings}  (a) shows PCA-projected embeddings of the SSL wav2vec 2.0~\cite{baevski2020wav2vec} model finetuned with the L2 loss. It can be observed how the representation learned by the layer attached to the output layer is fragmented and tends to cluster degradations while MOS is only partially projected along one of the 2 PCA dimensions. 
Other studies such as RnC~\cite{zha2024rank}, have observed this problem in non-audio regression minimising the L1 loss. When attaching a linear projection layer to representations that are ordered with respect to targets, RnC is able to improve all the baselines trained with the L1 loss across a wide range of regression tasks.

Motivated by the issues above, we present Speech COntrastive REgression for Quality (SCOREQ) assessment. We propose the SCOREQ loss function which is based on contrasting triplets to learn a quality manifold as shown in Figure \ref{fig:embeddings} (b). We adapt the batch-all triplet loss for person re-identification~\cite{ding2015deep,hermans2017defense} which addresses a classification task. Since MOS is a continuous label, the batch-all strategy is not trivial. By assuming that closeness is relative in regression, our loss function generates on-the-fly masks that identify valid triplets based on the MOS and are able to contrast all the possible triplet combinations in the batch based on regression targets. In addition, we investigate an alternative version that replaces the fixed margin with an adaptive margin initially proposed in~\cite{ha2021deep}. This study suggests that supervised contrastive learning for regression performs significantly better than L2 loss minimisation for speech quality assessment. 
Our contributions are (i) We conduct a systematic evaluation to prove that domain mismatch i.e. a lack of robustness to both out-of-distribution (ODS) and out-of-domain (ODM) test sets, is a serious problem in speech quality metrics; (ii) We propose a \textit{triplet loss function for contrastive regression} that is based on on-the-fly triplet mining and adaptive margin; (iii) We conduct an extensive statistical analysis on the results showing that our contrastive loss improves significantly results on both ODS and ODM data using two architectures and 11 test sets from several speech domains; (iv) We propose 2 ready-to-use speech quality metrics based on the SCOREQ loss for natural and synthetic speech that both work in two modes: no-reference (NR SCOREQ) and non-matching reference (NMR SCOREQ) mode i.e., where quality is measured with unpaired clean speech.

\section{Related Work}
\paragraph{Speech Quality Assessment} The earliest no-reference speech quality methods are based on signal processing~\cite{malfait2006p,kim2007anique+} but showed poor generalisation. Data-driven no-reference metrics improved results. However, often they are trained with quality labels obtained from full-reference speech quality metrics~\cite{kumar2023torchaudio,catellier2020wawenets,Quality-Net} which already suffer from generalisation issues. Other MOS predictors were trained on MOS scores~\cite{reddy2021dnsmos,mittag21_interspeech,avila2019non} and showed improvement across several databases. Recent methods are based on finetuning pre-trained self-supervised learning (SSL) models~\cite{tseng21b_interspeech,becerra22_interspeech,huang22f_interspeech} such as HuBERT~\cite{hsu2021hubert} or wav2vec 2.0~\cite{baevski2020wav2vec}. Although large improvement has been observed, e.g., in out-of-domain languages~\cite{becerra22_interspeech} we show in this paper that SSL finetuning still suffers from domain mismatch. 
Other metrics consider predicting speech quality without relying on the MOS. For example, SpeechLMScore~\cite{maiti} uses a speech-language model, and VQScore~\cite{fu2024selfsupervised} uses vector quantization from clean speech only. The NORESQA framework~\cite{manocha2021noresqa,manocha22c_interspeech} and the NOMAD model~\cite{ragano2024nomad} also propose solutions without MOS and they address the relative nature of quality assessment by using non-matching references. However, as we show in our experiments, both models still suffer from domain generalisation and overfitting of their corresponding training data similar to the no-reference models. 
\vspace{-2mm}
\paragraph{NOMAD vs. SCOREQ} 
NOMAD is trained by minimising the contrastive loss in a regression fashion, using the Neurogram Similarity Index Measure (NSIM)~\cite{hines2012speech} as a proxy for quality. SCOREQ shares the same principle as NOMAD but enhances both training efficiency and performance. A significant drawback of NOMAD is the offline preparation of triplets, which is inefficient because not all triplets are contrasted with each other, leading to suboptimal results. Additionally, offline triplet requires defining \textit{hard} and \textit{easy} triplets either by measuring the embedded distance at each epoch or by using labels. NOMAD employs the latter method, using the NSIM space to establish harder triplets. However, the NSIM does not always align with human perception.
In this paper, we replicate the NOMAD approach using MOS and demonstrate that it leads to suboptimal results, particularly in out-of-domain evaluations. To address this issue, SCOREQ is trained by learning all valid triplet combinations within the batch and removing the easy triplets. One more issue is that, because NSIM is a relative similarity criterion unlike MOS, triplets are composed only of the same clean reference. This approach is not optimal for learning a quality manifold where data are ordered based on quality, regardless of speech content. Therefore, we replace NSIM with MOS, allowing us to compare any combination of speech samples.
Our proposed adapted triplet loss resolves all these issues simultaneously, yielding better results without NOMAD's overhead of preparing triplets offline. While NOMAD does not require human labels such as MOS for training, we use MOS labels to demonstrate that the SCOREQ loss improves generalisation performance and robustness over the L2 loss when MOS labels are available. Unlike NOMAD which targets unsupervised quality prediction, our objective is to present a supervised quality metric as well as to show how the SCOREQ loss improves generalisation and robustness.
\paragraph{Contrastive Learning}
Contrastive learning is a widely adopted technique across various fields, such as computer vision, natural language processing, and audio processing, for self-supervised representation learning~\cite{chen2020simple,tian2020makes,le2020contrastive,PCL}. In the speech domain, general-purpose speech representation models based on contrastive learning have emerged to tackle tasks like speech recognition and speaker classification with minimal supervision~\cite{baevski2020wav2vec,mohamed2022self,milde18_interspeech,oord2018representation,jiang21_interspeech,chung2021w2v}. This technique is typically framed as a classification task, where categorical labels are defined.
Contrastive learning has also been applied to quality prediction, though it remains within a classification framework. For example, in CDPAM, content and acoustic representations are disentangled~\cite{manocha2021cdpam}. Beyond representation learning, some studies have employed contrastive learning for supervised classification tasks. SupCon~\cite{khosla2020supervised} demonstrates contrastive learning for image classification, proving more robust than the cross-entropy loss in various scenarios~\cite{kang2020exploring,li2022selective,zeng2021modeling}.

However, there has been less focus on using contrastive learning for regression tasks. Unlike classification, contrastive regression approaches are primarily supervised, using label distance to determine feature similarity~\cite{ordinalregression,dhaini2024contrastive,wang2022contrastive}. Rank-and-Contrast (RnC)~\cite{zha2024rank} is a supervised contrastive learning framework designed for regression problems. Although it is supervised, RnC uses label distance as a pre-training step for feature extraction. Their work suggests that speech quality prediction could benefit from such a representation, given the continuous nature of MOS.
Research has shown that contrastive learning pre-training often outperforms end-to-end supervised learning across several domains~\cite{karthik2021tradeoffs} and is more robust to data corruptions~\cite{zhong2022self}, making it a suitable approach for addressing domain mismatch issues~\cite{wang2022contrastive}. Our SCOREQ loss contributes not only to the speech quality domain but also to the relatively unexplored area of contrastive learning for regression.
\section{Contrastive Regression for MOS prediction}
We consider a training dataset $\mathbb{S}=\{\bm{x}_i, y_i\}$ where $\bm{x}_i \in \mathbb{X}$ are raw representations of the $i$-th speech sample and $y_i \in \mathbb{Y}$ is the corresponding MOS label such that a distance $r(\bm{x}_j,\bm{x}_k)=|y_j-y_k|$ can be defined between any two samples in the set $\mathbb{S}$. 

Let us consider an encoder $g:\mathbb{X}\mapsto \mathbb{H}$, that maps raw audio $\mathbb{X}$ to representations $\mathbb{H}$ and a projection head $f:\mathbb{H}\mapsto \mathbb{Z}$ that reduces the representations to a low-dimensional embedding space. Given 3 random speech files defined as anchor $\bm{x}_a$, positive $\bm{x}_p$, and negative $\bm{x}_n$ our goal is to learn a continuous-aware feature space where the distance between embeddings $D$ is ordered based on the similarity criteria $r(\bm{x}_j,\bm{x}_k)$:

\begin{equation}
\label{eq:1}
    r(\bm{x}_a, \bm{x}_p) < r(\bm{x}_a, \bm{x}_n) \Rightarrow D(f(g(\bm{x}_a)), f(g(\bm{x}_p))) < D(f(g(\bm{x}_a)), f(g(\bm{x}_n))) 
\end{equation}

Since we define the similarity criteria $r(\cdot,\cdot)$ as the MOS distance, equation \ref{eq:1} becomes:

\begin{equation}
\label{eq:2}
    |y_a - y_p| < |y_a - y_n| \Rightarrow D(f(g(\bm{x}_a)), f(g(\bm{x}_p))) < D(f(g(\bm{x}_a)), f(g(\bm{x}_n))) 
\end{equation}

\paragraph{Triplet Loss for Regression} To induce the defined order in the embedded space \(\mathbb{Z}\), we propose a triplet loss inspired by the classification framework but adapted for regression.
Given a training batch of size \(N\), the classification triplet loss is defined as~\cite{schroff2015facenet}:
\begin{equation} \label{eq:triplet_loss}
 \sum_{i}^{N} \max \left(0, ||f(g(\bm{x}_a^i)) - f(g(\bm{x}_p^i))||_2^2 - ||f(g(\bm{x}_a^i)) - f(g(\bm{x}_n^i))||_2^2 + m\right)
\end{equation}

The objective of this loss function is to maximise the similarity in the embedded space between the positive sample \(\bm{x}_p\) and the anchor \(\bm{x}_a\), while minimising the similarity between the negative sample \(\bm{x}_n\) and the anchor. The margin \(m\) is a constant that defines the distance threshold between the anchor and the negative sample, beyond which training stops. The classification triplet loss requires categorical labels such that \(y^a = y^p\) and \(y^a \neq y^n\).  However, the triplet loss can be adapted for regression by considering that the concept of closeness is relative in regression problems. For example, if the MOS of the anchor is 2.0, this should be closer to MOS 1.5 (positive) than to MOS 3.0 (negative). This concept has been successfully applied by the NOMAD speech quality metric~\cite{ragano2024nomad} and by the RnC framework with a different contrastive loss function~\cite{zha2024rank}. Nonetheless, as discussed earlier, NOMAD does not leverage the triplet loss efficiently.
\paragraph{Batch All Strategy for Regression}
The SCOREQ loss adapts the batch-all (BA) strategy used in~\cite{ding2015deep,hermans2017defense} for regression. To achieve this, we compute masks for each batch using the available MOS labels, assuming that every sample can serve as an anchor with potential valid triplets. The masks are generated with the understanding that in regression problems, the concept of closeness is relative, unlike the absolute nature of classification settings. After finding the valid triplets, we also remove the easy triplets i.e., the ones where the positive is closer to the anchor than the negative in the embedding space.

Given a training batch of size \(N\), for every possible distinct triplet combination in the batch \((i, j, k)\), we compute a 3D mask \(\mathcal{M}(i, j, k)\) such that:

\begin{equation}
  \mathcal{M}(i,j,k)=
  \begin{cases}
    1, & \text{if}\ (|y_i - y_j| < |y_i - y_k|) \wedge (i \neq j \neq k) \\
    0, & \text{otherwise}
  \end{cases}
\end{equation}

The SCOREQ loss for one batch is defined as:

\begin{equation}
  \label{eq:fixed}
  \mathcal{L} = \sum_{i=1}^N \sum_{j=1}^N \sum_{k=1}^N \mathcal{M}(i,j,k) \cdot \left( ||f(g(\bm{x}_i)) - f(g(\bm{x}_j))||_2 - ||f(g(\bm{x}_i)) - f(g(\bm{x}_k))||_2 + m \right)
\end{equation}

\begin{figure}[t]
\centering
\includegraphics[width=0.90\textwidth]{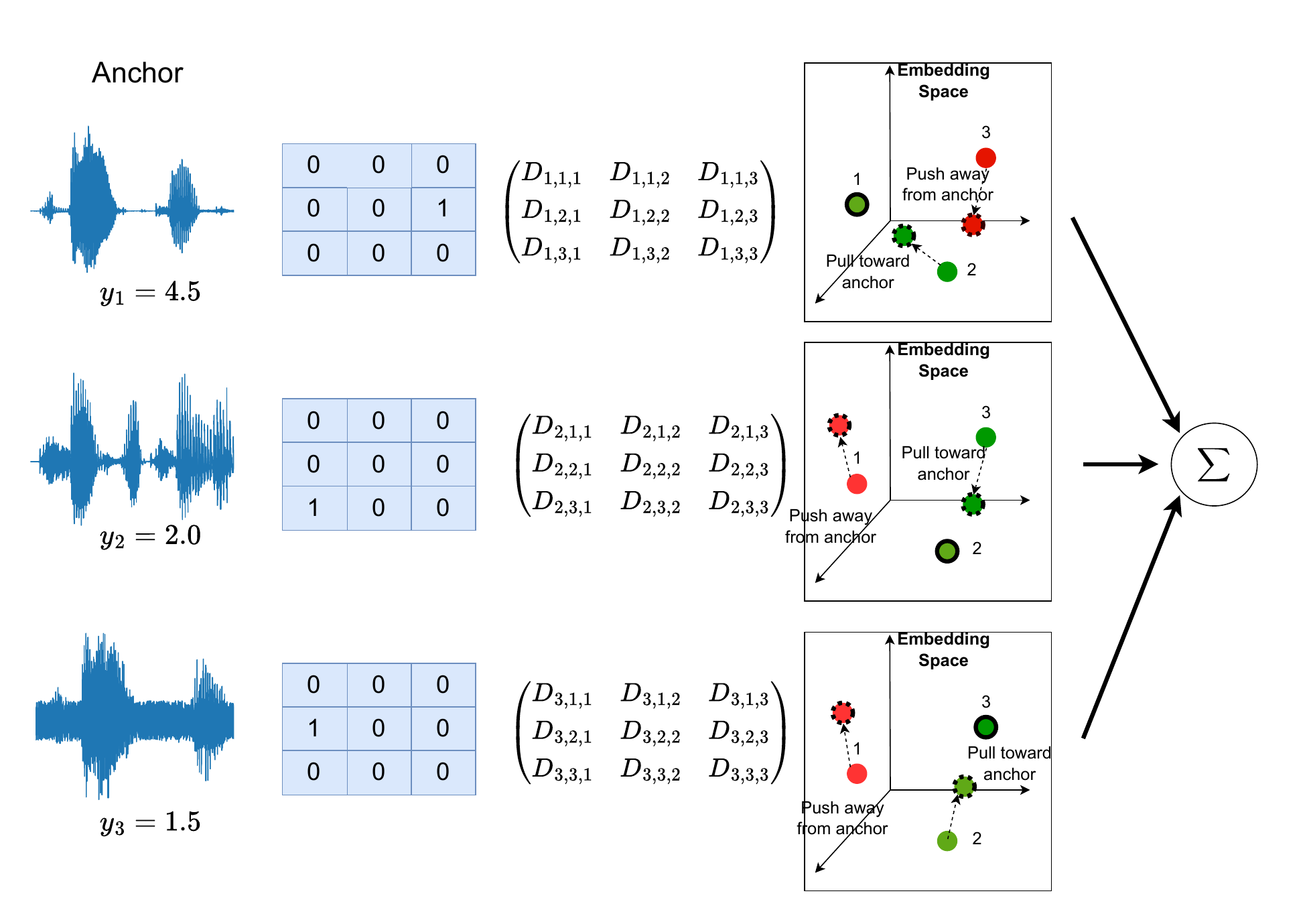}
\caption{Example of the SCOREQ loss using 3 samples in a training batch with corresponding MOS labels 4.5, 2.0, and 1.5. The distance matrix entries are defined as $D_{i,j,k} = ||f(g(\bm{x}_i))-f(g(\bm{x}_j))||_2 < ||f(g(\bm{x}_i)) - f(g(\bm{x}_k))||_2$. The intuition behind this contrastive loss for regression is shown in how the negative embeddings change in the anchor sample 1 where MOS is 4.5. We observe that the negative (sample 3 with MOS 1.5 ) will be further from the anchor with respect to sample 2. Indeed, because of the anchor 2 loss (where MOS is 2.0), sample 3 embeddings are pushed towards sample 2.}
\label{fig:batch_all}
\end{figure}

\begin{figure}[!t]
\centering
\includegraphics[width=0.80\textwidth]{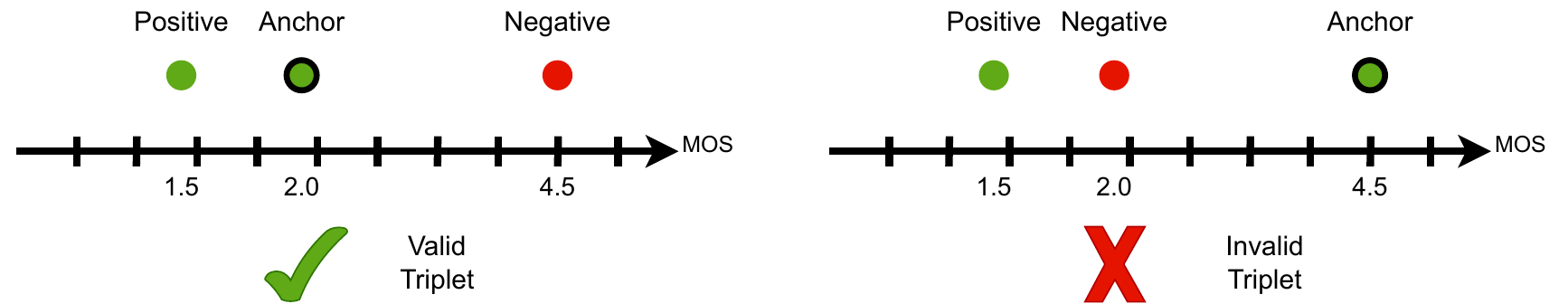}
\caption{Example of how the mask $M(i,j,k)$ assigns 0 or 1. If the distance between the anchor and positive is lower than the distance between anchor and negative we consider it as a valid triplet. The inequality condition must also be verified i.e., $i \neq j \neq k$}
\label{fig:valid_triplets}
\end{figure}

By multiplying by the mask \(\mathcal{M}(i,j,k)\), we ensure that only triplets where the MOS distance between the anchor and the positive is smaller than the MOS distance between the anchor and the negative are used. Figure \ref{fig:batch_all} illustrates an example of the SCOREQ loss function using a batch size of 3. This approach enables the model to compare every possible triplet in the batch in a relative manner, reflecting the human capacity to judge quality by using relative anchors for sensory judgement~\cite{lawless2010sensory}. The SCOREQ loss allows for training quality assessment models by sampling random data in the batch and fully utilising fast broadcasting computations. Figure \ref{fig:valid_triplets} shows examples of valid and invalid triplets, indicating that the validity of training on a specific triplet combination depends on the relative distances concerning the anchor. The mask \(\mathcal{M}(i,j,k)\) ensures that the training process ignores the invalid triplets.

\paragraph{Adaptive Margin}
Since we are addressing a regression problem, using a fixed margin $m$ is not optimal. In classification problems, the margin is typically set to a constant to ensure that training stops once the distance in the embedding space between the anchor and the negative exceeds $m$. However, given our continuous label space, we can leverage the labels to model an adaptive triplet loss. An adaptive margin for the triplet loss was initially proposed in \citet{ha2021deep}. In our paper, we explore combining the adaptive margin from~\cite{ha2021deep} with our batch all strategy loss for regression. 
Our experiments demonstrate that the significant performance improvement stems primarily from adopting the batch-all strategy over the offline triplet sampling used in NOMAD. The additional benefit of using an adaptive margin compared to a constant margin is minimal.

The idea behind the adaptive triplet loss is to replace the fixed margin \(m\) with a margin based on the MOS distance, as follows:
\begin{equation}
    \label{eq:adaptive}
    \begin{aligned}        
    \mathcal{L} = \sum_{i=1}^N\sum_{j=1}^N\sum_{k=1}^N \mathcal{M}(i,j,k) \cdot ( ||f(g(\bm{x}_i)) - f(g(\bm{x}_j)) ||_2 - ||f(g(\bm{x}_i)) - f(g(\bm{x}_k)) ||_2 + \\ (|y_i - y_j| - |y_i - y_k|)/(N-1))
    \end{aligned}
\end{equation}

The adaptive margin is divided by \(N-1\) to normalise it within the range [0,1]~\cite{ha2021deep}. 
The adaptive margin allows training to continue based on the MOS distance. This indicates that training will only stop if the distance between the negative sample and the anchor exceeds at least the normalised MOS distance. This is confirmed by computing the gradient (Appendix \ref{appendix:gradient}).

\begin{figure}[!t]
\centering
\includegraphics[width=0.90\linewidth]{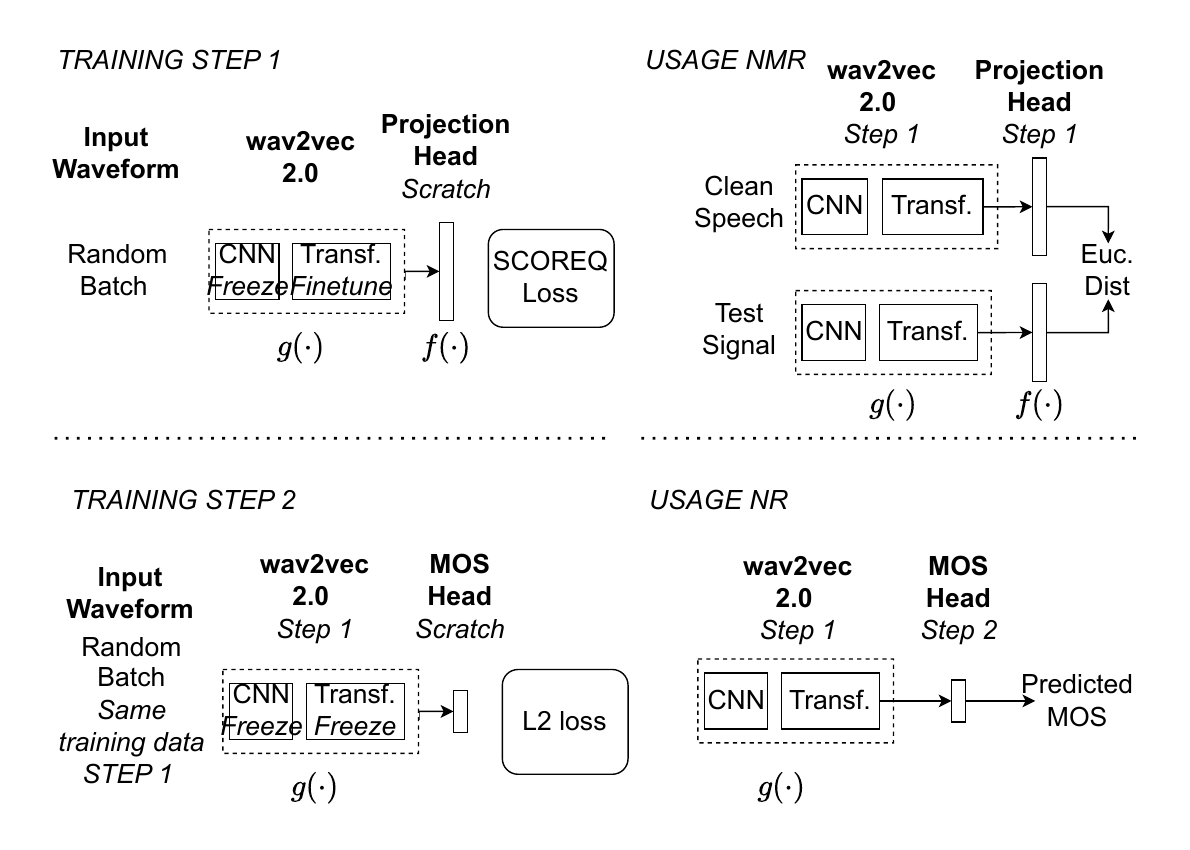}
\caption{SCOREQ modes. No-Reference (NR) mode is trained in 2 steps. We first pre-train the encoder $g(\cdot)$ with the SCOREQ loss. Next, we learn a linear layer (MOS head) that predicts an interpretable numerical MOS.}
\label{fig:scoreq_omdes}
\end{figure}

\paragraph{SCOREQ Training}
The SCOREQ model has the flexibility to be used in non-reference (NR) mode to predict interpretable MOS ratings and in non-matching reference (NMR) mode to predict a distance from unpaired clean speech. The latter has the advantage of being used as a perceptual training loss~\cite{ragano2024nomad,manocha2021cdpam} since SCOREQ is differentiable. 
For both approaches, we train an encoder $g:\mathbb{X}\mapsto \mathbb{H}$ that extracts representations $h_i=g(\bm{x}_i)$ and a projection head (embedding layer) that maps the representations to the space where the SCOREQ loss is applied: $f:\mathbb{H}\mapsto \mathbb{Z}$ producing $d$-dimensional embeddings $z_i=f(\bm{h}_i)$. For the NMR mode, we do not do further training. When evaluating we take random clean speech and measure the distance in the embedding space as done in NOMAD~\cite{ragano2024nomad} as follows: $D(\bm{x}_{test}, \bm{x}_{nmr}) = || f(g(\bm{x}_{test})) - f(g(\bm{x}_{nmr})) ||_2 $ where $\bm{x}_{test}$ is the degraded signal we want to predict quality for, $\bm{x}_{nmr}$ is a non-matching reference clean speech. 
For the NR-mode we freeze the encoder $g(\cdot)$ trained with SCOREQ and we train a linear layer that maps the embeddings $\bm{h}_i$ to MOS using the L2 loss. Since the encoder is frozen, the L2 loss does not affect the learned representation $\bm{h}_i$ which only depends on the SCOREQ loss. Our NR-mode model is compared with the L2 loss baseline which trains directly the same encoder $g(\cdot)$ using the L2 loss. In the baseline, the learned feature representation $\bm{h}_i$ depends only on the L2 loss minimisation. Notice that in the SCOREQ NR mode, we remove the projection head $f(\cdot)$ as done in other pre-trained contrastive learning approaches~\cite{manocha2021cdpam,chen2020simple}. This also makes sure that the L2 loss baseline architecture is the same as the NR mode SCOREQ architecture. 

\paragraph{Architecture}
We train the SCOREQ loss models with 2 architectures. The first is based on finetuning wav2vec 2.0 (w2v)~\cite{baevski2020wav2vec} \textit{BASE} model as done in NOMAD. The other architecture (w2vlight) is a smaller version of w2v where the transformer network is reduced from 12 to 4 layers and trained from scratch. We evaluate w2vlight to study whether without using the pre-trained transformer and a less powerful architecture we obtain the same effect with SCOREQ. More details in Appendix \ref{appendix:architecture}.
Similar to NOMAD, for SCOREQ the encoder output $g(\bm{x}_i)$ is the average of the last transformer layer along the time dimension. The embedding layer $f(\cdot)$ is a small network consisting of a ReLU followed by a fully connected layer projecting the encoder representations to a 256-dimensional feature vector. In all experiments, we only finetune the transformer of w2v, leaving the CNN frozen as done in other studies~\cite{ragano2024nomad}. The best model is found with early stopping on a validation set. More details are in Appendix \ref{appendix:training_details}.

\begin{table}[!b]
  \centering
  \caption{Metrics and corresponding training datasets}
  \fontsize{8pt}{8pt}\selectfont
    \begin{tabularx}{\textwidth}{llllllll}
    \hline
    \textbf{Metric} & \textbf{DB name} & \textbf{Dom. ID} &\textbf{Domain}  & \textbf{BW (kHz)} \\
    \hline
    NISQA & NISQA TRAIN SIM~\cite{mittag21_interspeech}   & D1 & Sim Degradations & 48 \\

    NR-PESQ, NR-SI SDR & DNS Squim~\cite{kumar2023torchaudio} & D2 & DL Speech Enhancement &  16\\

   NORESQA-MOS &  VoiceMOS Train~\cite{huang22f_interspeech} & D3 & Speech Synthesis &   16 \\

    \hline
    \end{tabularx}
  \label{tab:train_sets}
\end{table}

\section{Generalisation of Speech Quality Metrics}
In this section, we demonstrate that certain no-reference speech quality models do not generalise well. Table \ref{tab:train_sets} lists the metrics and their corresponding training sets used in our experiments. For each quality metric, we compare in-domain (IN), out-of-distribution (ODS), and out-of-domain (ODM) performances. We define ODS as test sets that are within the same domain as the training set but exhibit a distribution shift. For example, with simulated telephone degradations, a set consisting of real phone calls or a subset of degradations is out-of-distribution but within the same training domain. Similarly, a set in another language with the same degradations is considered a distribution shift but remains within the same domain. In contrast, we label test sets designed for other application domains as ODM. For example, speech synthesis is ODM with respect to telephone speech training data. Detailed information about training and test sets used is provided in the Appendix \ref{appendix:trainsets} and \ref{appendix:testsets} respectively.
We selected three domains for our experiments: (i) We train the NISQA metric on the NISQA TRAIN SIM dataset~\cite{mittag21_interspeech} that includes simulated degradations for telephone speech; (2) The no-reference (NR) versions of PESQ and SI-SDR provided in \textit{TorchAudio-Squim}~\cite{kumar2023torchaudio} that are trained on deep learning-based speech enhancement artifacts and noisy speech; (3) NORESQA-MOS~\cite{manocha22c_interspeech} which was trained on the train partition of the VoiceMOS challenge dataset~\cite{huang22f_interspeech} that includes text-to-speech and voice conversion speech. We use the version provided in \textit{TorchAudio-Squim}~\cite{kumar2023torchaudio}.

\begin{table}[!t]
\caption{Domain mismatch evaluated with Pearson's correlation (PC). Each metric performs the best in its corresponding training domain. We indicate the domain shift between training and test with IN\like{10}, ODS\like{5}, ODM\like{0}. Our proposed metrics show the best generalisation performance across most of the datasets. Note that results on the dataset DNS Squim are taken from~\cite{kumar2023torchaudio} since the dataset is not publicly available.}
\centering
\Huge
\ra{1.1}
\begin{adjustbox}{max width=0.99\textwidth}
\fontsize{22pt}{22pt}\selectfont
\begin{tabular}{@{}lcccccccccccccccccccccccccccccccccccccccccccccccccc@{}}
   
\cmidrule{1-8}
\multicolumn{2}{c} {\phantom} & 
\multicolumn{1}{c} {\textbf{NISQA}} & \multicolumn{1}{c}{\textbf{NR-PESQ}} & \multicolumn{1}{c}{\textbf{NR-SI SDR}} & \multicolumn{1}{c}{\textbf{NORESQA-M}} & \multicolumn{1}{c}{\textbf{NR-SCOREQ}} (\textit{Ours}) \\

\cmidrule(r){3-3} \cmidrule(r){4-5} \cmidrule(r){6-6}  \cmidrule{7-7}
\textbf{Dataset} & &  D1 & \multicolumn{2}{c}{D2} &  D3 & \multicolumn{1}{c}{D1} \\

\cmidrule{2-8}

NISQA TEST FOR  & D1\like{10}D2\like{5}D3\like{0} & 0.91 & 0.79 & 0.74 & 0.68  &  \textbf{0.97}   \\
NISQA TEST P501 & & 0.94  & 0.88   & 0.81  & 0.70  &  \textbf{0.96}  \\
\cmidrule{2-8}

DNS Squim   & D2\like{10} & //  & \textbf{0.96} & \textbf{0.99} & //  & //  & \\
\cmidrule{2-8}

VoiceMOS Test 1   & D1\like{0}D2\like{0}D3\like{10} & 0.54   & 0.71  & 0.67  & 0.85   & \textbf{0.86}   \\
VoiceMOS Test 2   & & 0.64  & 0.49  & 0.55  & \textbf{0.91}  &  0.82   \\
\cmidrule{2-8}

NOIZEUS        & D1\like{5}D2\like{5}D3\like{0} & 0.85  & 0.75  & 0.70  & 0.15  &  \textbf{0.91} \\
NISQA TEST LT  & & 0.84  & 0.66 & 0.56 & 0.60 & \textbf{0.86}  \\
P23 EXP3       & & 0.82  & 0.77 & 0.17 & 0.71 & \textbf{0.94}  \\
TCD VOIP       & & 0.76  & 0.76 & 0.76 & 0.61 & \textbf{0.85} \\
TENCENT     & & 0.78  & 0.78 & 0.77 & 0.57 & \textbf{0.86}  \\
\cmidrule{2-8}

P23 EXP1       & D1\like{5}D2\like{0}D3\like{0} & 0.76  & 0.70  & 0.82  & 0.40  &  \textbf{0.96}     \\
\cmidrule{2-8}

TENCENT-Rev     & D1\like{0}D2\like{0}D3\like{0} & 0.40  & 0.36  & 0.32  & 0.36   &  \textbf{0.79}      \\

\bottomrule
\label{tab:mismatch}
\end{tabular}
\end{adjustbox}
\end{table}

\begin{figure}[!t]
\centering
\includegraphics[width=0.90\linewidth]{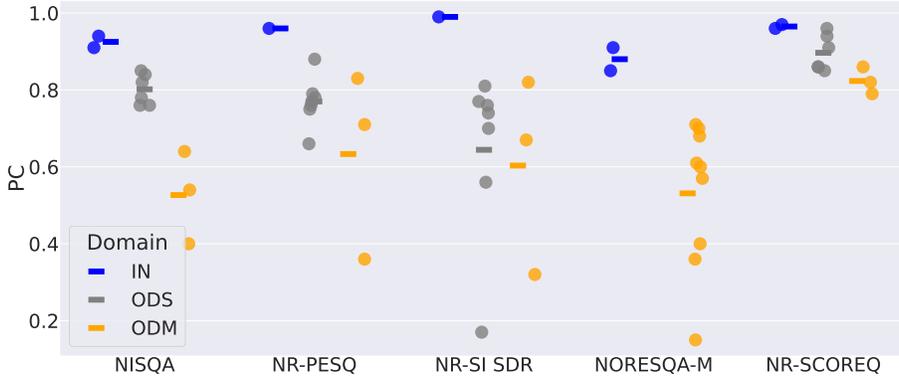}
\caption{Domain mismatch. Each dot is a dataset, while horizontal lines represent the PC average for each domain shift (IN, ODS, ODM). PC values are the same of Table \ref{tab:mismatch}.}
\label{fig:domain_mismatch}
\end{figure}


\subsection{Performance Evaluation}
In Table \ref{tab:mismatch} and Figure \ref{fig:domain_mismatch}, we report the results on the test datasets using the Pearson Correlation Coefficient (PC) to measure the linear relationship between predictions and MOS ground truth. Notice that each test set is ODS or ODM relative to the training set domains D1 (simulated telephone speech), D2 (deep learning-based speech enhancement), and D3 (speech synthesis). We also include one of our SCOREQ loss models in no-reference (NR-SCOREQ) mode. Our proposed method is outlined in the below sections. It is included in Table \ref{tab:mismatch} to show the improvement over  ODM and ODS domains. 
 
Comprehensive results, including Spearman's rank correlation (SC) and root mean squared error (RMSE) are similar to PC performance (Appendix \ref{appendix:performance_eval}). Our results confirm the hypothesis that speech quality metrics show difficulties in reaching satisfactory results in new domains (ODM) or even within the same domain but with a distribution shift (ODS). 

\section{Experiments}
Experiments were conducted to compare L2 loss training against SCOREQ loss and evaluate performance across three scenarios: IN, ODS, and ODM. We assessed two domains: telephone speech degradations using the NISQA TRAIN SIM dataset, and speech synthesis using the VoiceMOS training partition. 

\subsection{Simulated Degradations: Models}
\paragraph{Offline Model} This model replicates the approach used in the NOMAD speech quality metric but with MOS as a label. We prepared 46783 triplets from the NISQA TRAIN SIM dataset, which were chosen using the "hard" approach reported in the NOMAD paper that improved results. In this approach, for every sampled anchor we take the 10 most difficult triplets based on the MOS i.e., the ones where the MOS distance between anchor-positive difference and the anchor-negative difference is the shortest. 
\paragraph{SCOREQ Models} We evaluate two variants: SCOREQ \textit{Const} and SCOREQ \textit{Adapt} consisting of using the SCOREQ loss with fixed margin (Eq. \ref{eq:fixed}) and adaptive margin (Eq. \ref{eq:adaptive}) respectively. 
\paragraph{L2 loss} The L2 loss model is the baseline and consists of finetuning the wav2vec 2.0 transformer by attaching an output layer that minimises the L2 loss between MOS targets and predictions.  
\paragraph{NOMAD} We evaluate this model since is based on a similar principle and uses the same SCOREQ architecture. The original model only comes in NMR mode. However, we train an NR mode with the same SCOREQ approach. We freeze the NOMAD model and train an output layer to map NOMAD representations to an interpretable MOS scale. This is the only NR model where step 1 is trained without using MOS since we use the NOMAD pre-trained model that is unsupervised.

\subsection{Simulated Degradations: Results}
We evaluate the linearity (PC) and the average precision (RMSE) for NR metrics. For NMR metrics we still use linearity (PC) but also the Spearman's coefficient (SC) since NMR models do not predict MOS magnitude but they learn how to rank based on quality. Results for the RMSE are reported after performing a one-degree polynomial mapping between predictions and ground truth to adjust for the listening test bias in each dataset, as it is typically recommended in quality metric evaluation. NMR performance is calculated by extracting 50 random speech files from the test-clean partition of Librispeech~\cite{panayotov2015librispeech}. These are used in all the NMR models found in this paper. 
\begin{table}[!h]
\caption[C]{Performance evaluation using PC($\uparrow$), RMSE($\downarrow$) for NR metrics and PC and SC($\uparrow$) for NMR metrics. Domain shift (IN, ODS, ODM) is labelled with respect to the training set NISQA TRAIN SIM.}
\centering
\Huge
\ra{1.2}
\begin{adjustbox}{max width=\textwidth}
\fontsize{30pt}{30pt}\selectfont
\begin{tabular}{@{}lccccccccccccccccccccccccccccccccccccccccccccccccccccccccccc@{}}
\toprule
\multicolumn{2}{c} {\phantom} & \multicolumn{12}{c} {\textit{NR}} & \multicolumn{2}{c} {\phantom}  & \multicolumn{10}{c}{\textit{NMR}}  \\

\cmidrule(r){3-17} \cmidrule{19-26}

\multicolumn{2}{c} {\phantom} & \multicolumn{3}{c} {\textbf{L2 loss}} & \multicolumn{3}{c} {\textbf{NOMAD}} & \multicolumn{3}{c} {\textbf{Offline}} &  \multicolumn{6}{c} {\textbf{SCOREQ} \textit{Ours}}   & \multicolumn{3}{r}{\textbf{NOMAD}}  & \multicolumn{6}{c}{\textbf{SCOREQ} \textit{Ours}}  \\

\cmidrule(r){4-5} \cmidrule(r){7-8} \cmidrule(r){10-11}  \cmidrule(r){13-17} \cmidrule{19-20} \cmidrule{22-26} 
\multicolumn{2}{c} {\phantom} & \multicolumn{10}{c} {\phantom} & \multicolumn{2}{c}{\textit{Const}} &\multicolumn{1}{c} {\phantom} & \multicolumn{2}{c}{\textit{Adapt}} & \multicolumn{4}{c} {\phantom} &  \multicolumn{2}{c}{\textit{Const}} & \multicolumn{1}{c} {\phantom}  & \multicolumn{2}{c}{\textit{Adapt}}\\

\textbf{Test Set} & Domain &&  PC & RMSE &&  PC & RMSE && PC & RMSE && PC & RMSE && PC & RMSE && PC & SC && PC & SC && PC & SC
\\ \cmidrule{1-26}

NISQA TEST FOR    & IN && 0.96 & 0.23 && 0.89 & 0.38  && 0.89 & 0.38  && 0.96 & 0.22 && \textbf{0.97} & \textbf{0.21}  && 0.73 & 0.60 && \textbf{0.97 }& \textbf{0.95} && 0.96 & \textbf{0.95} \\

NISQA TEST P501   & IN && 0.95 & 0.29 && 0.91 & 0.40  && 0.88 & 0.46  && \textbf{0.96} & \textbf{0.29} && \textbf{0.96} & \textbf{0.28}     && 0.79 & 0.76  && 0.93 & 0.95 &&  0.92 & \textbf{0.96}    \\

\hline
 
P23 EXP1    & ODS && 0.95 & 0.25   && 0.93 & 0.27  && 0.85  & 0.41 && \textbf{0.96} &\textbf{ 0.21} && \textbf{0.96} & 0.22    && 0.89 & 0.90 &&\textbf{ 0.97} & \textbf{0.97 }&& 0.96 & 0.96     \\

P23 EXP3    & ODS && \textbf{0.95} & \textbf{0.23}   && 0.92 & 0.28  && 0.86 & 0.37 && 0.92 & 0.29 && 0.94 & 0.25   && 0.83 & 0.70 && 0.90 & \textbf{0.81} && 0.87 & 0.80   \\

NOIZEUS & ODS && 0.84 & 0.26  && 0.72 & 0.33  && \textbf{0.92} & \textbf{0.19}  && 0.90 & 0.21 && 0.91 & 0.20  && 0.47 & 0.42 && 0.91 & \textbf{0.91} && 0.87 & 0.87   \\

NISQA TEST LIVETALK & ODS && 0.85 & 0.43  && 0.84 & 0.44  && 0.79 & 0.50  && 0.87 & \textbf{0.40} && 0.86 & 0.42 && 0.79 & 0.73 && 0.85 & 0.85 && \textbf{0.88} & \textbf{0.88}   \\

TCD VOIP   &   ODS && 0.80 & 0.59 && 0.80 & 0.58  && 0.78 & 0.60 &&\textbf{ 0.85} &\textbf{ 0.51} && \textbf{0.85} & \textbf{0.51}  && 0.65  & 0.57 && 0.80 & \textbf{0.66} && 0.74 & \textbf{0.66} &  \\

TENCENT     & ODS && 0.81 & 0.71  && 0.80 & 0.72   && 0.82 & 0.69 && 0.84 & 0.66 && \textbf{0.86} & \textbf{0.62}    && 0.67 & 0.76 && 0.80 & \textbf{0.86} && 0.85 &\textbf{ 0.86} &       \\

\hline

TENCENT-Rev     & ODM && 0.74 & 0.61  && 0.71 & 0.64   && 0.69 & 0.65 && 0.78 & \textbf{0.57} && \textbf{0.79} & \textbf{0.56}  && 0.57 &  0.40  && 0.73 & 0.74 && 0.78 & \textbf{0.76} \\


VoiceMOS Test 1    & ODM && 0.75 & 0.53  && 0.79  & 0.49   && 0.81 & 0.47 && 0.84 & 0.43 && \textbf{0.86} & \textbf{0.41}      && 0.74 & 0.71 && 0.78 & 0.76 && 0.74 &\textbf{ 0.77}  \\

VoiceMOS Test 2    & ODM && 0.69 & 0.68  && 0.67  & 0.70   && 0.57 & 0.77 && 0.75 & 0.62 && \textbf{0.82} & \textbf{0.53}     && 0.42 & 0.50 && 0.74 & \textbf{0.76} && 0.75 & 0.75  \\

\cmidrule{1-26}
\label{tab:simdegr_results}
\end{tabular}
\end{adjustbox}
\end{table}

Table \ref{tab:simdegr_results} illustrates the results for IN, ODS, and ODM scenarios. We observe that both SCOREQ \textit{Const} and SCOREQ \textit{Adapt} are the best for both some ODS data and all ODM data. On average, our proposed loss contributes to improving the performance of no-reference speech quality prediction. The \textit{Offline} model does not perform as well as our, showing inferior results than the L2 loss. We also evaluated the \textit{Offline} model encoder representations in NMR mode (Appendix \ref{appendix:nmr_performance}), showing that it does not capture quality as well as SCOREQ. In Table \ref{tab:w2vlight} (Appendix) we show that the same improvement trend is confirmed with the w2vlight architecture. This suggests that our loss function can help training more compact networks without a significant performance drop. We notice that NOMAD underperforms SCOREQ but it has been trained with NSIM labels~\cite{hines2012speech} used as a proxy for MOS quality.

\subsection{Speech Synthesis: Results}
The same approach is evaluated using the VoiceMOS training sets~\cite{huang22f_interspeech} to examine the generalisation from speech synthesis to telephone speech degradations. We report results from NORESQA-MOS which has been trained on the same sets. Again, we compare the SCOREQ loss models against the L2 loss. We do not evaluate the \textit{Offline} model since it did not perform well in the simulated degradations domain.  
Results in Table \ref{tab:voicemos} show that the SCOREQ representations improve results over the L2 loss. Also, the NMR SCOREQ models significantly improve ODM performances over NORESQA-MOS. No difference can be found for IN sets (Table \ref{tab:simdegr_results}).

\begin{table}[!t]
\caption[C]{Performance evaluation using PC($\uparrow$), RMSE($\downarrow$) for NR metrics and PC and SC($\uparrow$) for NMR metrics. Domain shift (IN, ODM) is labelled with respect to the training set VoiceMOS Train.}
\centering
\Huge
\ra{1.3}
\begin{adjustbox}{max width=0.95\textwidth}
\fontsize{20pt}{20pt}\selectfont
\begin{tabular}{@{}lccccccccccccccccccccccccccccccccccccccccccccccccccccccccccc@{}}
\toprule
\multicolumn{2}{c} {\phantom} & \multicolumn{9}{c} {\textit{NR}}   & \multicolumn{9}{c}{\textit{NMR}}  \\

\cmidrule(r){3-11} \cmidrule{12-20}

\multicolumn{2}{c} {\phantom} & \multicolumn{3}{c} {\textbf{L2 loss}}  &  \multicolumn{6}{c} {\textbf{SCOREQ} \textit{Ours}}  & \multicolumn{3}{c}{\textbf{NORESQA-M}}  & \multicolumn{6}{c}{\textbf{SCOREQ} \textit{Ours}}  \\

\cmidrule(r){4-5} \cmidrule(r){6-12}  \cmidrule(r){13-14} \cmidrule{15-20} 
\multicolumn{2}{c} {\phantom} & \multicolumn{4}{c} {\phantom} & \multicolumn{2}{c}{\textit{Const}} &\multicolumn{1}{c} {\phantom} & \multicolumn{2}{c}{\textit{Adapt}} & \multicolumn{4}{c} {\phantom} &  \multicolumn{2}{c}{\textit{Const}} & \multicolumn{1}{c} {\phantom}  & \multicolumn{2}{c}{\textit{Adapt}}\\

\textbf{Test Set} & Domain &&  PC & RMSE && PC & RMSE && PC & RMSE && PC & SC && PC & SC && PC & SC
\\ \cmidrule{1-20}

VoiceMOS Test 1    & IN && 0.90 & 0.36 &&  0.90 & 0.35 && \textbf{0.91} & \textbf{0.34} && 0.85 & \textbf{0.87} && 0.87 & 0.86 && 0.83 & 0.80       \\

VoiceMOS Test 2    & IN && \textbf{0.98} & \textbf{0.18}   && 0.97 & 0.21 && 0.97 & 0.21 && 0.91 & \textbf{0.90} && 0.93 & 0.85 && 0.88 & 0.83        \\

\cmidrule{1-20}

NISQA TEST FOR    & ODM && 0.82 & 0.47  && 0.85 & 0.44 && \textbf{0.87} & \textbf{0.41} && 0.68 & 0.58 && 0.85 & 0.78 && 0.84 & \textbf{0.79}  \\

NISQA TEST P501   & ODM && 0.86 & 0.50    && \textbf{0.89} & \textbf{0.45}  && \textbf{0.89} & \textbf{0.45} && 0.70 & 0.81 && 0.86 & 0.87 && 0.84 & \textbf{0.88}  \\

P23 EXP1    & ODM && 0.92 & 0.31    && \textbf{0.93}  & \textbf{0.29} && 0.92 & 0.30 && 0.40 & 0.35 && 0.84 & 0.91 && 0.93 & \textbf{0.95} \\

P23 EXP3    & ODM && 0.85 & 0.38    && 0.88 & \textbf{0.35} && 0.87 & 0.36 && 0.71 & 0.48 && 0.83 & 0.69 && 0.90 & \textbf{0.85 }    \\

NOIZEUS & ODM && 0.59 & 0.39  && 0.69 & \textbf{0.35}  && 0.67 & 0.36 && 0.15 & 0.15 && 0.48 & 0.46 && \textbf{0.63} & \textbf{0.60}\\

NISQA TEST LIVETALK & ODM && 0.81 & 0.47  && 0.83 & 0.46  && 0.84 & \textbf{0.44} && 0.60 & 0.51 && 0.78 & 0.74  && \textbf{0.83} & \textbf{0.80}    \\

TCD VOIP   &   ODM && 0.83 & 0.54   && 0.87 & \textbf{0.48} && 0.86 & 0.49 && 0.61 & 0.71 && 0.81 & \textbf{0.81} && 0.77 & 0.61   \\

TENCENT    & ODM && 0.78 & \textbf{0.76}  && 0.76 & 0.78 &&  0.77 & 0.78 && 0.57 & 0.59  && \textbf{0.71} & 0.78 && 0.78 & 0.79    \\

TENCENT-Rev     & ODM && 0.35 & 0.85   && 0.43 & 0.82 && 0.44 & \textbf{0.81} && 0.36 & 0.20 && 0.48 & \textbf{0.33} && \textbf{0.49} & 0.38       \\

\cmidrule{1-20}
\label{tab:voicemos}
\end{tabular}
\end{adjustbox}
\end{table}

\subsection{Statistical Analysis and Encoder Representation Performance}
We conduct a bootstrap procedure to understand whether the SCOREQ correlation improvement observed above is due to chance. Our results show that the SCOREQ models perform significantly better than the NR L2 loss in most of the ODM and ODS sets (Appendix \ref{appendix:statistical_analysis}). We compare the encoder representations $g(\cdot)$ in NMR mode to understand which one explains quality information the best, showing that L2 loss encoders perform significantly worse than encoders trained with SCOREQ (Appendix \ref{appendix:nmr_performance}).

\section{Conclusions}
This paper presents SCOREQ, a contrastive regression loss to learn continuous representations for speech quality metrics. We have demonstrated that substituting the classic L2 loss with our proposed SCOREQ loss results in consistent performance improvements across 11 datasets and 2 speech domains. The SCOREQ loss can encourage a new research direction to train speech quality metrics. By learning ordered representations that follow the continuous nature of MOS, SCOREQ improves generalisation. In addition, we made available the SCOREQ metric in both no-reference mode and non-matching reference (NMR) mode for both simulated speech degradations and speech synthesis. The NMR models have been shown to be better than other NMR metrics. They can be used as a perceptual training loss and serve better applications where the reference is available (e.g., speech codecs) or quality wants to be measured relative. Finally, we believe that the SCOREQ loss can be applied to any application where models can be trained with regression-based targets.

\section{Acknowledgements}
This publication has emanated from research conducted with the financial support of Science Foundation Ireland under Grant numbers 12/RC/2289\_P2 and 13/RC/2106\_P2 and a gift from Google. For the purpose of Open Access, the author has applied a CC BY public copyright licence to any Author Accepted Manuscript version arising from this submission.

\bibliographystyle{plainnat}
\bibliography{ref}

\newpage

\appendix
\renewcommand{\appendixpagename}{APPENDIX}
\begin{center}  
\appendixpagename
\end{center}

\section{Software}
Every model is trained on an Nvidia Tesla V100 GPU. We trained our models in PyTorch~\cite{paszke2019pytorch} and Torchaudio~\cite{yang2022torchaudio}. The wav2vec 2.0 model is taken from the fairseq toolkit~\cite{ott2019fairseq}. The repository associated with this paper can be found here \url{https://github.com/alessandroragano/scoreq}.

\section{Limitations}
\label{appendix:limitations}
In this paper and its appendix, we present an exhaustive list of experiments. However, it is crucial to address certain limitations that pave the way for future research, especially in understanding the scope of our proposed SCOREQ loss.
\paragraph{Paper Scope} We believe the SCOREQ loss holds potential for various regression tasks. This belief is underpinned by the motivation to address generalisation issues in speech quality assessment. Despite this, domain mismatch remains a challenge across numerous applications. A significant limitation of our study is that we did not evaluate the loss outside the speech domain, as this was beyond the scope of our current investigation.

\paragraph{Execution Time} We computed both L2 and SCOREQ loss using the same batch of data, loading 4 waveforms and measuring the computational time on the GPU for the entire operation required for one batch, including time to load waveforms, forward pass, loss calculation, backward pass, and zero gradients. On average, we found that the L2 loss takes ~0.65 seconds, while the SCOREQ loss takes ~0.75 seconds using our GPUs. This slight increase is expected since the batch-all strategy needs to compute distances between all triplet combinations, including invalid triplets which are discarded with the 3D boolean mask.

\paragraph{Cross-domain between encoder and target}
We have assessed two speech domains: simulated degradations on natural speech and synthetic speech. However, one scenario we have not explored is the performance of the model when the encoder is trained on one domain (step 1 in NR SCOREQ training) and the MOS mapping is trained on a different domain (step 2 in NR SCOREQ training).

\section{Impact}
\label{appendix:impact}
Assessing speech quality is critically important, as it helps avoid the expensive and time-consuming process of conducting listening tests. Given the rapid advancements in generative AI research within the speech domain—including text-to-speech, voice conversion, and neural speech coding—it is increasingly challenging to apply quality metrics that were developed for other domains such as old speech codecs. Understanding domain mismatches and working towards a generalisable speech quality metric could fundamentally assist researchers in generative AI. Moreover, contrastive regression remains a relatively unexplored approach. Through this paper, we aim to contribute to the development of improved regression solutions.  

\section{Training and Validation Sets}
\label{appendix:trainsets}
\paragraph{NISQA TRAIN SIM} The NISQA TRAIN SIM dataset~\cite{mittag21_interspeech} is part of the NISQA Corpus\footnote{NISQA Corpus: \url{https://github.com/gabrielmittag/NISQA/wiki/NISQA-Corpus}} and includes several degradations for telephone speech combined and isolated. Additive white Gaussian noise, signal correlated MNRU noise, randomly sampled noise clips taken from the DNS-Challenge dataset, lowpass, highpass, and bandpass filters with random cutoff frequencies, amplitude clipping, speech level changes, codecs in various bitrate modes, packet-loss conditions with random and bursty patterns. 
The dataset is made of 10000 speech samples, each degraded with a different condition. Source samples are in English and taken from the Librivox clips of the DNS challenge~\cite{reddy20_interspeech}, the TSP database~\cite{kabal2002tsp}, the UK and Ireland English dialect data~\cite{demirsahin2020open}, and AusTalk~\cite{burnham2011building} which includes interviews from Australian English. Overall, the dataset is composed of 2322 speakers. The sampling rate is 48000 kHz and MOS labels are obtained from 5 listeners on average. 

\paragraph{NISQA VAL SIM} The NISQA VAL SIM dataset~\cite{mittag21_interspeech} is a partition made with the same conditions and source samples of NISQA TRAIN SIM. This dataset consists of 2500 speech samples with 938 different speakers. We use this dataset for early stopping in all the experiments and to find the best combination of learning rate and batch size of the L2 loss model. 

\paragraph{DNS Squim} This dataset was used to train the NR-PESQ and NR-SI SDR models that are available in \textit{TorchAudio-Squim}~\cite{kumar2023torchaudio}. The dataset is made of deep learning-based speech enhancement models (GCRN architecture~\cite{tan2019learning}) and noisy speech trained on the DNS challenge dataset~\cite{reddy20_interspeech}. Training and test partitions are made of 364500 and 22800 speech samples respectively. This dataset is labelled with full-reference metric scores from PESQ wideband~\cite{pesqwb} and SI SDR~\cite{le2019sdr}.  

\paragraph{VoiceMOS Train} The training partition of the VoiceMOS challenge~\cite{huang22f_interspeech} includes several text-to-speech and voice conversion samples from the Blizzard and the Voice Conversion challenges~\cite{cooper21_ssw}. The VoiceMOS challenge dataset is labelled with 8 listeners for each clip. The training partition includes 175 synthesis systems. We include this dataset since it has been used to train the NORESQA-MOS~\cite{manocha22c_interspeech} metric available in \textit{TorchAudio-Squim}. The challenge includes two training partitions, main and OOD tracks. We combine both together to train SCOREQ in the speech synthesis domain.

\section{Test Sets}
\label{appendix:testsets}
In all experiments, we use 11 test sets labelled with MOS obtained from several speech domains and languages. Details for each dataset are shown in Table \ref{tab:test_sets}. 

\begin{table}[!h]
  \centering
  \caption{Test datasets}
  \fontsize{9pt}{9pt}\selectfont
    \begin{tabularx}{\textwidth}{lXXXXXX}
    \hline
    \textbf{DB name} & \textbf{Domain} & \textbf{Cond} & \textbf{Samples} & \textbf{Lang} & \textbf{BW(kHz)} \\
    \hline
    NISQA TEST P501~\cite{mittag21_interspeech}  & Sim Degradations & 60 & 240 & en & 48 \\
    
    NISQA TEST FOR~\cite{mittag21_interspeech}& Sim Degradations & 60 & 240 & en  & 48  \\

    NISQA TEST LIVETALK~\cite{mittag21_interspeech} & Real Phone Calls & 58 & 232 & de  & 48  \\

    P23 EXP 1~\cite{ITUT1998}  & Codecs & 50 & 200 & en  & 16  \\

    P23 EXP 3~\cite{ITUT1998} & Codec, Bursty patterns, Noise & 50 & 200 & en  & 16  \\

    
    TCD VOIP~\cite{harte2015tcd}  & Noise, Chop, Clip, Comp. Speakers, Echo & 80 & 320 & en & 48      \\

    NOIZEUS~\cite{hu2006subjective}  & Sig. Proc. Speech Enhancement & 96 & 1536 &  en & 8      \\

    TENCENT-Rev~\cite{yi22b_interspeech} & Real-world Reverberation & 3197 & 3197 &  cn & 16      \\

    TENCENT~\cite{yi22b_interspeech}  & Sim Degradations & 8336 & 8366 &  cn & 16      \\

    VoiceMOS Test 1~\cite{huang22f_interspeech}   & Text-to-speech, voice conversion & 187 & 1066 &  en & 16      \\

    VoiceMOS Test 2~\cite{huang22f_interspeech} & Text-to-speech, voice conversion & 24 & 540 & en,jp,cn & 16 \\

    \hline
    \end{tabularx}
  \label{tab:test_sets}
\end{table}

All the test sets are used in their original version except for TCD VOIP where we remove the 4 MNRU conditions since they are anchors and also included in the NISQA TRAIN SIM dataset. By doing that we test a more difficult scenario.  

\section{Reproducing State of the Art Quality Metric Performances}
\label{appendix:sota}
\paragraph{NISQA}
\label{nisqa_training}
We train NISQA with the default parameters, including fullband mode (48000 kHz), and we use the NISQA early stopping approach with patience of 20 epochs on the validation set NISQA VAL SIM to find the best model. Depending on the bandwidth of the test set we only change the parameter \textit{ms\_fmax} that determines the maximum frequency to use to calculate the mel spectrograms. 
\begin{table}[!h]
\caption{NISQA \textit{ms\_fmax} values chosen based on the input bandwidth.}
\centering
\begin{tabular}{cc}
\toprule
 \textbf{Bandwidth Hz} & \textbf{\textit{ms\_fmax} Hz} \\
\midrule
  48000  &  20000   \\
  16000  &  8000   \\
  8000  &  4000   \\

\bottomrule
\end{tabular}
\label{tab:nisqa_parameters}
\end{table}
See Table \ref{tab:nisqa_parameters} for details. We do not resample the test files since the NISQA architecture adapts to any input sampling rate. Notice that the NISQA metric provided in the NISQA repo has been trained on multiple datasets and in a multi-task fashion predicting several quality dimensions beyond MOS. In our experiments, we are only interested in examining how each metric performs when trained in one domain only, so that we can run controlled experiments. This explains why the results reported in the NISQA paper~\cite{mittag21_interspeech} are different from ours.

\paragraph{\textit{TorchAudio-Squim} Results}
For the three metrics NR-PESQ, NR-SI SDR, and NORESQA-MOS we follow the tutorial provided in the Torch documentation\footnote{\href{https://pytorch.org/audio/main/tutorials/squim_tutorial.html}{\textit{TorchAudio-Squim}}}. 
To predict NORESQA-MOS we extract 50 random clean speech samples from the test-clean partition of the Librispeech dataset~\cite{panayotov2015librispeech}. 
All the test files are resampled to 16 kHz which is the expected sampling rate from all 3 metrics. 

\section{SCOREQ Architectures}
\label{appendix:architecture}
\paragraph{wav2vec 2.0} SCOREQ architecture is based on finetuning the transformer of the self-supervised learning (SSL) wav2vec 2.0 model. We use the \textit{BASE} version. This model has been pre-trained to learn general-purpose representations using a contrastive learning approach. The architecture is made of 7 1D convolutional neural network (CNN) layers that encode audio frames of 25 ms into 400-dimensional feature vectors. Each frame-wise CNN output is fed into a transformer made of 12 layers that outputs $T$ 768-dimensional context vectors. Positional encoding for the transformer is learned with a convolutional neural network. We take the average of the $T$ frames and attach the output of the average, which is one vector of dimension $768$ to an embedding layer that is made of a ReLU followed by a linear layer of dimension $256$. This represents the dimension of the final embeddings. This approach was used by the NOMAD metric~\cite{ragano2024nomad}. 
\paragraph{w2vlight}
We train the same approach with a smaller architecture. The w2vlight is built by reducing the number of the transformer layers from 12 to 4 and finetuning the transformer from scratch. We keep the same dimensionality and the same positional encoding approach of the original wav2vec 2.0 architecture as well as the number of neurons of the feedforward neural network that composes each transformer layer. 

\section{Training Details}
\label{appendix:training_details}
\paragraph{SCOREQ loss}
Training is done by trimming all the input files to 4 seconds which is enough to capture quality and avoids using large memory. All the results on the test data are computed without trimming, using the original length for each file.
We finetune the transformer with a learning rate set to $0.00001$, while the embedding layer trained from scratch uses a learning rate of $0.001$. We use the Adam optimizer with default PyTorch settings except for the learning rate. The batch size is set to $128$ in all SCOREQ experiments. Both learning rates decay exponentially with a decay factor of $0.99$ every $10$ epochs without improvement. 
In all the SCOREQ experiments, training is stopped if the Spearman correlation coefficient (SC)  on the NISQA VAL SIM dataset does not improve for more than 100 epochs. The best model is taken as the one with the highest SC. SC is calculated in NMR mode by extracting 200 random clean speech files from the dev-clean partition of the Librispeech dataset~\cite{panayotov2015librispeech}.

\paragraph{L2 loss}
The models with the L2 loss are trained with the same architectural configurations of the SCOREQ loss models. The only difference is in the output layer. Instead of the embedding layer, the average of the transformer output is attached to a linear layer with one output neuron to predict MOS. Input files are also trimmed to 4 seconds to have a fair comparison while the original waveform length is used in all the test sets. Unlike SCOREQ, for the L2 loss we perform a gridsearch using several combinations of learning rate and batch size to select the best model on the validation set NISQA VAL SIM. This is done to avoid any potential result difference only due to different hyperparameters. In all the experiments, training is stopped if the loss function on the NISQA VAL SIM dataset does not improve for more than 100 epochs. The best model is taken as the one with the lowest L2 validation loss. We did some informal experiments by stopping with the best SC (as in SCOREQ) but we did not notice any significant difference in the results. Notice that even step 2 of NR-SCOREQ models has been validated using the L2 loss as a validation metric and not with the SC.

\begin{table}[!h]
\caption{Hyperparameter search for the L2 loss model.}
\centering
\begin{tabular}{ccc}
\toprule
 \textbf{Batch Size} & \textbf{Learning Rate} & \textbf{Validation Loss} \\
\midrule
  32  &  0.1 & 0.2970 \\
  64  &  0.1 &  0.2865\\
  128  &  0.1 &  0.2913\\
  32  &  0.01 & 0.2505\\
  64  &  0.01 &  \textbf{0.2466}\\
  128  &  0.01 & 0.2730  \\
\bottomrule
\end{tabular}
\label{tab:gridsearch_l2loss}
\end{table}

Table \ref{tab:gridsearch_l2loss} shows that the best configuration is batch size 64 and learning rate 0.01. We use the Adam optimizer with default PyTorch settings except for the learning rate. Notice that these parameters have been found using the original wav2vec 2.0 architecture but we also used them for the w2vlight. We observe that our proposed models based on the SCOREQ loss are not trained with an equivalent preliminary gridsearch. This means that better results could be found for SCOREQ. 

\section{Embedding Visualisation - Figure 1}
\label{appendix:figure1}
Figure \ref{fig:embeddings} shows the embeddings of the L2 loss model against our proposed SCOREQ loss. We use the models trained on NISQA TRAIN SIM and compute embeddings on TCD VOIP~\cite{harte2015tcd}.
Embedding vectors are projected onto 2D space using PCA. We project the encoder $g(\cdot)$ (i.e., the last transformer layer) of the L2 loss model trained on the original wav2vec 2.0 architecture. This is the layer that is attached to the output. For the SCOREQ-based model, we project the SCOREQ \textit{Const} version since it represents the novelty of our work. In this case, we project the output of the embedding layer $f(\cdot)$ which is the input of the SCOREQ loss. Notice that in Table \ref{tab:nmrmode} we compute MOS correlation performance using the encoder representations $g(\cdot)$ demonstrating that quality is predicted significantly better than the L2 loss also in the encoder representations and not only in the embedding layer. The Pearson's correlation (PC) results reported in the Figure are calculated in NMR mode, they correspond to the same results reported in Table \ref{tab:simdegr_results} for the SCOREQ \textit{Const} model. The PC of the L2 loss is taken from Table \ref{tab:nmrmode}. PC is computed to show the amount of quality information that is explained in both feature representations. The Normalised Mutual Information (NMI) scores are computed after running K-Means with 5 clusters using the same representations of PC calculation. The NMI is computed between the output of the K-Means algorithm and the 5 degradation labels. This is done to quantify the amount of degradations explained by both feature representations. The results show that the L2 loss feature representation tends to cluster degradation information more than SCOREQ loss. Instead, our proposed loss function learns ordered representations with respect to target MOS. Although degradation information is linked to quality it is also orthogonal i.e., two degradation conditions are labelled with equal MOS. So having representations with lower NMI between cluster embeddings and degradations is preferred for MOS quality prediction.  

\section{Gradient Adaptive Margin}
\label{appendix:gradient}
The usefulness of the adaptive margin is confirmed by inspecting the gradient. Assuming that the anchor is \(i\), the positive is \(j\), and the negative is \(k\), the gradient with respect to the anchor sample is:
\begin{equation}
  \frac{\partial \mathcal{L}}{\partial \bm{x}_i} =
  \begin{cases}
    \bm{x}_k - \bm{x}_j, & \text{if}\  ||\bm{x}_i - \bm{x}_j||_2 + \frac{|y_i - y_j| - |y_i - y_k|}{N-1} > ||\bm{x}_i - \bm{x}_k||_2 \\
    0, & \text{otherwise}
  \end{cases}
\end{equation}
The gradient for the positive and the negative samples can be obtained with the same approach. 

\section{Experiments using w2vlight architecture}
\label{appendix:w2vlight}
We run the same experiments using the NISQA TRAIN SIM corpus but with the w2vlight architecture. Table \ref{tab:w2vlight} shows that the SCOREQ loss improves over the L2 loss in every test set. This means that less powerful architectures could also benefit from our proposed approach.

\begin{table}[!h] 
\caption{Pearson's Correlation (PC) performance using w2vlight. We use the SCOREQ \textit{Adapt} model, which was the best with the original wav2vec 2.0 architecture. }
\fontsize{8pt}{8pt}\selectfont
\begin{minipage}{0.5\textwidth}
\begin{tabular}{lcc}
\toprule
\textbf{Dataset} & \textbf{L2 loss} & \textbf{SCOREQ} (\textit{Ours}) \\
\midrule
NISQA TEST FOR  &  0.83  &  \textbf{0.86}  \\
NISQA TEST P501  &  0.86  &  \textbf{0.90}   \\
P23 EXP1  &  \textbf{0.85}  &  \textbf{0.85}   \\
P23 EXP3  &  0.85  &  \textbf{0.92}   \\
NOIZEUS  &  0.71  &  \textbf{0.80}   \\

\bottomrule
\end{tabular}

\end{minipage} \hfill
\begin{minipage}{0.5\textwidth}
\begin{tabular}{lcc}
\toprule
\textbf{Dataset} & \textbf{L2 loss} & \textbf{SCOREQ} (\textit{Ours}) \\
\midrule
NISQA TEST LIVETALK  &  0.66  &  \textbf{0.80}   \\
TCD-VOIP  &  0.85  &  \textbf{0.87}   \\
TENCENT  &  0.80  &  \textbf{0.83}   \\
TENCENT-Rev &  0.25  &  \textbf{0.35}  \\
VoiceMOS Test 1  &  0.69  &  \textbf{0.75}   \\
VoiceMOS Test 2 &  \textbf{0.75}  &  0.69  \\

\bottomrule
\end{tabular}

\end{minipage}
\label{tab:w2vlight}
\end{table}

\begin{table}[!b]
\caption{Performance evaluation with Spearman's Rank Correlation Coefficient (SC).}
\centering
\Huge
\ra{1.1}
\begin{adjustbox}{max width=\textwidth}
\fontsize{22pt}{22pt}\selectfont
\begin{tabular}{@{}lcccccccccccccccccccccccccccccccccccccccccccccccccc@{}}
   
\cmidrule{1-8}
\multicolumn{2}{c} {\phantom} & 
\multicolumn{1}{c} {\textbf{NISQA}} & \multicolumn{1}{c}{\textbf{NR-PESQ}} & \multicolumn{1}{c}{\textbf{NR-SI SDR}} & \multicolumn{1}{c}{\textbf{NORESQA-M}} & \multicolumn{1}{c}{\textbf{NR-SCOREQ}} (\textit{Ours}) \\

\cmidrule(r){3-3} \cmidrule(r){4-5} \cmidrule(r){6-6}  \cmidrule{7-7}
\textbf{Dataset} & &  D1 & \multicolumn{2}{c}{D2} &  D3 & \multicolumn{1}{c}{D1} \\

\cmidrule{2-8}

NISQA TEST FOR  & D1\like{10}D2\like{5}D3\like{0} & 0.89 & 0.73 & 0.74 & 0.58  &  \textbf{0.95}  \\
NISQA TEST P501 & & 0.94  & 0.88   & 0.81  & 0.81  &  \textbf{0.97} \\
\cmidrule{2-8}

VoiceMOS Test 1   & D1\like{0}D2\like{0}D3\like{10} & 0.50   & 0.71  & 0.65  & \textbf{0.87}   &  \textbf{0.87} & \\
VoiceMOS Test 2   & & 0.35  & 0.51  & 0.62  & \textbf{0.90}    &  0.81  \\
\cmidrule{2-8}

NOIZEUS        & D1\like{5}D2\like{5}D3\like{0} & 0.86  & 0.72  & 0.66  & 0.15  &  \textbf{0.90}   \\
NISQA TEST LT  & & 0.78  & 0.59 & 0.58 & 0.51 & \textbf{0.85}   \\
P23 EXP3       & & 0.74  & 0.62 & 0.04 & 0.48 & \textbf{0.89}   \\
TCD VOIP       & & 0.81  & 0.79 & \textbf{0.87} & 0.71 & \textbf{0.85 } \\
TENCENT     & & 0.77  & 0.80 & 0.76 & 0.59 & \textbf{0.87}   \\
\cmidrule{2-8}

P23 EXP1       & D1\like{5}D2\like{0}D3\like{0} & 0.70  & 0.84  & 0.83  & 0.35  &  \textbf{0.96}   \\
\cmidrule{2-8}

TENCENT-Rev     & D1\like{0}D2\like{0}D3\like{0} & 0.39  & 0.16  & 0.30  & 0.20   &  \textbf{0.78}   \\

\bottomrule
\label{tab:mismatch_spearman}
\end{tabular}
\end{adjustbox}
\end{table}

\begin{table}[!h]
\caption{Performance evaluation root mean squared error (RMSE). }
\centering
\Huge
\ra{1.1}
\begin{adjustbox}{max width=\textwidth}
\fontsize{22pt}{22pt}\selectfont
\begin{tabular}{@{}lcccccccccccccccccccccccccccccccccccccccccccccccccc@{}}
   
\cmidrule{1-8}
\multicolumn{2}{c} {\phantom} & 
\multicolumn{1}{c} {\textbf{NISQA}} & \multicolumn{1}{c}{\textbf{NR-PESQ}} & \multicolumn{1}{c}{\textbf{NR-SI SDR}} & \multicolumn{1}{c}{\textbf{NORESQA-M}} & \multicolumn{1}{c}{\textbf{NR-SCOREQ}} (\textit{Ours}) \\

\cmidrule(r){3-3} \cmidrule(r){4-5} \cmidrule(r){6-6}  \cmidrule{7-7}
\textbf{Dataset} & &  D1 & \multicolumn{2}{c}{D2} &  D3 & \multicolumn{1}{c}{D1} \\

\cmidrule{2-8}

NISQA TEST FOR  & D1\like{10}D2\like{5}D3\like{0} & 0.34 & 0.50 & 0.55 & 0.60  &  \textbf{0.21}  \\
NISQA TEST P501 & & 0.32  & 0.47   & 0.56  & 0.69  &  \textbf{0.28}  \\
\cmidrule{2-8}

VoiceMOS Test 1   & D1\like{0}D2\like{0}D3\like{10} & 0.68   & 0.57  & 0.60  & 0.42   &  \textbf{0.41}  &  \\
VoiceMOS Test 2   & & 0.72 & 0.81  & 0.78  & \textbf{0.39}   &  0.53   \\
\cmidrule{2-8}

NOIZEUS        & D1\like{5}D2\like{5}D3\like{0} & 0.26  & 0.32  & 0.34  & 0.48  &  \textbf{0.20}  \\
NISQA TEST LT  & & 0.44  & 0.61 & 0.67 & 0.65 & \textbf{0.42}   \\
P23 EXP3       & & 0.41  & 0.46 & 0.71 & 0.51 & \textbf{0.25}   \\
TCD VOIP       & & 0.62  & 0.63 & 0.63 & 0.77 & \textbf{0.51}  \\
TENCENT     & & 0.50  & 0.42 & 0.44 & 0.70 & \textbf{0.62} \\
\cmidrule{2-8}

P23 EXP1       & D1\like{5}D2\like{0}D3\like{0} & 0.70  & 0.84  & 0.83  & 0.35  &  \textbf{0.22}   \\
\cmidrule{2-8}

TENCENT-Rev     & D1\like{0}D2\like{0}D3\like{0} & 0.39  & 0.16  & 0.30  & 0.20   &  \textbf{0.56}     \\

\bottomrule
\label{tab:mismatch_rmse}
\end{tabular}
\end{adjustbox}
\end{table}

\section{Evaluation of Ranking and Average Precision}
\label{appendix:performance_eval}
We compare the state-of-the-art quality metrics against our proposed models using Spearman's Rank Correlation Coefficient (SC) and the root mean squared error (RMSE) to evaluate the monotonic relationship and the average precision between predicted and ground truth MOS. In all our experiments we consider the predictions after one degree polynomial mapping to adjust for the bias of each listening test. This will only affect the RMSE values. 
We observe that each quality metric performs the best in test sets corresponding to the domain they are trained for. The NISQA metric's highest performance is reported for NISQA TEST P501 and NISQA TEST FOR which include several simulated degradations similar to the training set. Performances on other sets that are in the same domain but produced with different listening tests and unseen conditions are much lower (P23 EXP1, P23 EXP3, TCD VOIP).
The model does not equally perform for real phone calls (NISQA TEST LIVETALK) showing generalisation issues when training quality metrics with artificial degradations that try to emulate real phone calls. The significant gap between TENCENT-Rev and TENCENT suggests that quality metrics suffer when there are unseen conditions such as real-world reverberation (TENCENT-Rev), even if tested on distant languages (testing on Chinese and training on English). The no-reference PESQ and SI-SDR do not achieve the same results on their domain as reported in the Squim paper. In particular, they fail significantly for real phone calls and Chinese reverberated speech. NORESQA-MOS shows a performance drop in all the datasets that are not synthetic speech, showing a lower correlation for P23 EXP1 which includes speech encoded up to three times (transcoding). 

\section{Assessment of the Encoder $g(\cdot)$ representations.}
\label{appendix:nmr_performance}
As mentioned in the paper, the NR SCOREQ models are trained in 2 steps. We first train encoder $g(\cdot)$ and the embedding layer (projection head) $f(\cdot)$ with the SCOREQ loss. The second step is based on training a mapping between the encoder $g(\cdot)$ representations and MOS with the encoder frozen from step 1. We discard the embedding layer in this stage. 
Our main contribution is that encoder representations learned with the SCOREQ are not fragmented and allow easier mapping to MOS prediction than L2 end-to-end training. For this reason, here we analyse how much the encoder representations capture quality. To do that, we extract the output of the encoder $g(\cdot)$ and measure the Euclidean distance with non-matching reference as done in the NMR models illustrated in the paper. We evaluate linearity (PC) and ranking (SC) between the distance and the MOS labels. We use the same 50 unpaired clean speech samples from the Librispeech test-clean partition as non-matching references. We evaluate the encoder representations of the L2 loss baseline, NOMAD, the \textit{Offline} model, SCOREQ \textit{Const} and SCOREQ \textit{Adapt}. 

\begin{table}[!h]
\caption[C]{MOS correlations for encoder $g(\cdot)$ representation distances. Domain shift (IN, ODS, ODM) is labelled with respect to the training set NISQA TRAIN SIM.}
\centering
\Huge
\ra{1.1}
\begin{adjustbox}{max width=\textwidth}
\fontsize{16pt}{16pt}\selectfont
\begin{tabular}{@{}lcccccccccccccccccccccccccccccccccccccccccccccccccc@{}}\toprule

\multicolumn{2}{c} {\phantom} & \multicolumn{3}{c} {\textbf{L2 loss}} & \multicolumn{3}{c} {\textbf{NOMAD}} & \multicolumn{3}{c}{\textbf{Offline}} & \multicolumn{7}{c}{\textbf{SCOREQ} \textit{(Ours)}}  \\

\cmidrule{4-5} \cmidrule{7-8} \cmidrule{10-11} \cmidrule{13-18} 
 & && && && && && & \multicolumn{3}{c}{\textbf{\textit{Const}}} & \multicolumn{2}{c}{\textbf{\textit{Adapt}}}\\

\textbf{Test Set} & Domain &&  PC & SC && PC & SC && PC & SC && PC & SC && PC & SC 
\\ \midrule
NISQA TEST FOR    & IN && 0.75 & 0.75 && 0.71 & 0.65 && 0.59 & 0.54 && \textbf{0.94} & \textbf{0.92} && 0.86 & 0.85 &   \\

NISQA TEST P501   & IN && 0.76 & 0.85   &&  0.84 & 0.84  && 0.62 & 0.64  && \textbf{0.92} & \textbf{0.95} && 0.80 & 0.87    \\

\hline

P23 EXP1    & ODS && 0.89 & 0.90    && 0.88 & 0.88  && 0.57  & 0.62 && \textbf{0.97} & \textbf{0.96} && 0.93 & 0.94  \\

P23 EXP3    & ODS && 0.54 & 0.54   && 0.83 & 0.70  && 0.54 & 0.58 && \textbf{0.90} & \textbf{0.82} && 0.64 & 0.65     \\

NOIZEUS & ODS && 0.58 & 0.63  && 0.53 & 0.52  && 0.53 & 0.52  && \textbf{0.90} & \textbf{0.89} && 0.83 & 0.81   \\

NISQA TEST LIVETALK & ODS && 0.75 & 0.73 && 0.78 & 0.71  && 0.55 & 0.55  && \textbf{0.87} & \textbf{0.87} && 0.86 & 0.85   \\

TCD VOIP   &   ODS && 0.53 & 0.60 && 0.67 & 0.67  && 0.44 & 0.45 &&\textbf{ 0.82} & \textbf{0.85} && 0.57 & 0.50  \\

TENCENT    & ODS && 0.74 & 0.76 && 0.63 & 0.73 && 0.73 & 0.71 &&  \textbf{0.84} & \textbf{0.86} && \textbf{0.84} & 0.81    \\

\hline

TENCENT-Rev     & ODM && 0.60 & 0.59  && 0.55 & 0.43   && 0.54 & 0.55 && \textbf{0.80} & \textbf{0.78} && 0.75 & 0.76    \\

VoiceMOS Test 1    & ODM && 0.80 & 0.79 && 0.79  & 0.79   && 0.54 & 0.58 && \textbf{0.83} & \textbf{0.85} && 0.74 & 0.77      \\

VoiceMOS Test 2    & ODM && \textbf{0.76} & \textbf{0.78} && 0.35  & 0.45   && 0.45 & 0.38 && \textbf{0.80} & \textbf{0.79} && 0.73 & 0.71       \\

\bottomrule
\label{tab:nmrmode}
\end{tabular}
\end{adjustbox}
\end{table}

The results indicate that both SCOREQ models effectively capture quality information in the encoder output. We find that the L2 loss correlates less with MOS, suggesting that additional factors are entangled in the representation. This observation implies that our model could be trained on a large available MOS dataset to tune the encoder, and then adapt the linear layer using a smaller, domain-specific annotated dataset. This approach is particularly beneficial in domains where data is scarce, as collecting MOS through listening tests is often resource-intensive.

\section{Statistical Analysis}
\label{appendix:statistical_analysis}
We consider the case of overlapping and dependent correlation coefficients i.e. when one variable is in common (MOS ground truth in our case) and we want to test whether correlation $\rho(X_{MOS}, Y_{SCOREQ})$ is different from correlation $\rho(X_{MOS}, Y_{L2})$. For each test set $T$, we resample $N_T$ times with replacement, where $N_T$ is the sample size of dataset $T$. We compute PC using bootstrap samples and then we compute the difference $\rho_{d} = \rho(X_{MOS}, Y_{SCOREQ}) - \rho(X_{MOS}, Y_{L2}) $. We execute this procedure 15000 times in each dataset and use the empirical distribution to compute a 95\% confidence interval of the difference between the two correlation coefficients. We conclude that the 2 correlations are significantly different if the confidence interval does not include 0. We conduct a two-tail test and report the corresponding \textit{p}-values to evaluate significance. The null hypothesis is that the difference between the two correlations includes the zeros. If \textit{p}-value is lower than 0.05 we have evidence to reject this hypothesis and conclude that the two correlation coefficients are statistically different. We conduct a statistical difference between the NR SCOREQ \textit{Adapt} model against the L2 loss model trained on both NISQA TRAIN SIM (Table \ref{tab:simdegr_results}) and VoiceMOS Train \ref{tab:voicemos}. In this way, we test the generalisation of training on simulated degradations and testing on synthetic speech and vice-versa. These statistical tests examine the flexibility of SCOREQ that can be adapted to train models in more than one domain. 

\begin{figure}[!h]%
    \centering
    \subfloat[\centering TCD VOIP]{{\includegraphics[width=0.45\textwidth]{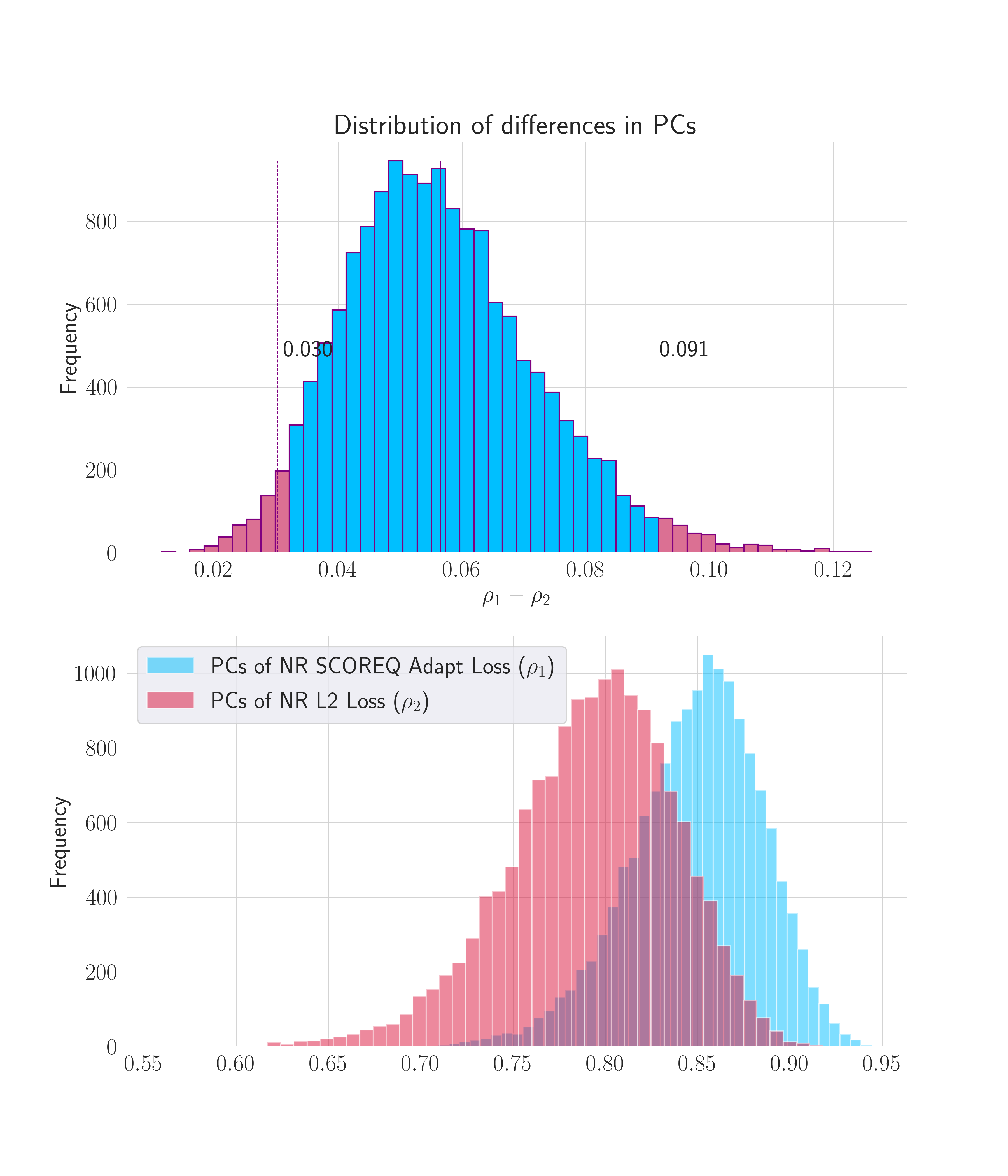} }}%
    \qquad
    \subfloat[\centering VoiceMOS Test 1]{{\includegraphics[width=0.45\textwidth]{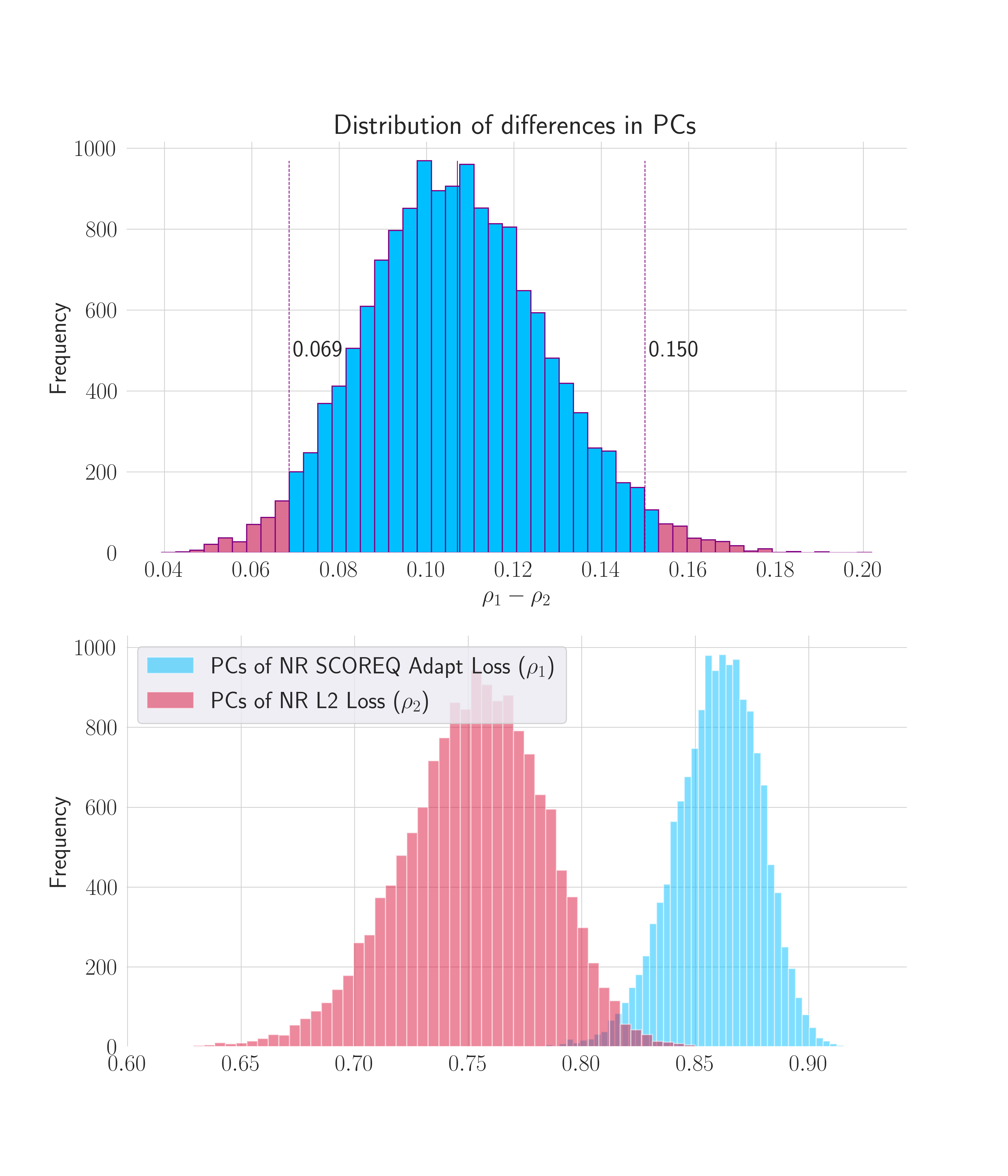} }}%
    \caption{Bootstrapping on 2 datasets where NR SCOREQ \textit{Adapt} performs better than the L2 loss. We can see how in many subsets, our model shows improvement. We only report these 2 datasets but the same trend is confirmed in all the test sets where SCOREQ is better as shown in Table \ref{tab:statistics}}.%
    \label{fig:example}%
\end{figure}

\paragraph{Simulated degradations}
Our statistical analysis in Table \ref{tab:statistics} finds that SCOREQ is significantly better than the L2 loss model in all the ODS and ODM datasets except for the NISQA LIVE TALK datasets. There is no difference for IN sets, which further confirms the generalisation issues of deep learning-based speech quality metrics. Figure \ref{fig:example} shows the bootstrapping procedure on 2 datasets.

\begin{table}[!h]
  \centering
  \caption{Bootstrap statistics of PC computed on NR SCOREQ \textit{Adapt} against L2 loss trained on the NISQA TRAIN SIM datasets. Domain shift (IN, ODS, ODM) is labelled relatively to the simulated degradations domain of the training set.}
  \fontsize{8.6pt}{8.6pt}\selectfont
    \begin{tabularx}{\textwidth}{lXXlXXX}
    \hline
    \textbf{Dataset} & \textbf{Domain} & \textbf{\textit{p}-value} & \textbf{Confidence Interval} & \textbf{Outcome Test} & \textbf{SCOREQ} (PC) & \textbf{L2} (PC) \\
    \hline
NISQA TEST FOR      & IN  & 0.185 & [-0.002, 0.014]     &	No Diff. & 0.97 & 0.96  \\
NISQA TEST P501     & IN  & 0.348 & [-0.004, 0.0012]    &   No Diff. & 0.96 & 0.95 \\
P23 EXP1            & ODS & 0.319 &	[-0.011, 0.033]	    &	No Diff. & 0.96 & 0.95 \\
P23 EXP3            & ODS & 0.325 & [-0.035, 0.009]     &   No Diff. & 0.92 & 0.95  \\
NOIZEUS             & ODS & 0.000 & [0.230, 0.444]      &   \textbf{SCOREQ}   & 0.90 & 0.84 \\ 
NISQA TEST LIVETALK & ODS & 0.450 & [-0.020, 0.037]	    &   No Diff. & 0.87 & 0.83 \\
TCD VOIP            & ODS & 0.000 & [0.030, 0.091]	    &	\textbf{SCOREQ}   & 0.85 & 0.80 \\
TENCENT             & ODS & 0.000 & [0.043, 0.052]      &   \textbf{SCOREQ}   & 0.86 & 0.81 \\
TENCENT-Rev         & ODM & 0.000 & [0.034, 0.055]	    &	\textbf{SCOREQ}   & 0.79 & 0.74 \\
VoiceMOS Test 1	    & ODM & 0.000 & [0.069, 0.150]	    &	\textbf{SCOREQ}   & 0.86 & 0.75 \\
VoiceMOS Test 2	    & ODM & 0.001 & [0.043, 0.248]	    &	\textbf{SCOREQ}   & 0.82 & 0.69 \\
    
    \hline
    \end{tabularx}
  \label{tab:statistics}
\end{table}

\paragraph{Speech synthesis}
We conduct same bootstrapping for the models trained on the speech synthesis domain. Table \ref{tab:stats_synth} shows that SCOREQ is significantly better than the L2 loss in 5 test sets worse in only one. This confirms that the SCOREQ loss can improve generalisation for speech quality metrics. 

\begin{table}[!h]
  \centering
  \caption{Bootstrap statistics of PC computed on NR SCOREQ \textit{Adapt} against L2 loss trained on the VoiceMOS Train dataset. Domain shift (IN, ODS, ODM) is labelled relatively to synthetic speech domain of the training set.}
  \fontsize{8.6pt}{8.6pt}\selectfont
    \begin{tabularx}{\textwidth}{lXXlXXX}
    \hline
    \textbf{Dataset} & \textbf{Domain} & \textbf{\textit{p}-value} & \textbf{Confidence Interval} & \textbf{Outcome Test} & \textbf{SCOREQ} (PC) & \textbf{L2} (PC) \\
    \hline
VoiceMOS Test 1	    & IN & 0.341 & [0.016, 0.096]	     & No Diff. & 0.91 & 0.90 \\
VoiceMOS Test 2	    & IN & 0.167 & [0.043, 0.248]	     & No Diff. & 0.97 & 0.98 \\
NISQA TEST FOR      & ODM  & 0.011 & [0.016, 0.096]      & \textbf{SCOREQ} & 0.87 & 0.82  \\
NISQA TEST P501     & ODM & 0.029 & [0.002, 0.057]       & \textbf{SCOREQ} & 0.89 & 0.86 \\
P23 EXP1            & ODM & 0.730 &	[-0.021, 0.030]	     & No Diff. & 0.92 & 0.92 \\
P23 EXP3            & ODM & 0.627 & [-0.052, 0.089]      & No Diff. & 0.87 & 0.85  \\
NOIZEUS             & ODM & 0.002 & [0.034, 0.145]       & \textbf{SCOREQ} & 0.67 & 0.59 \\ 
NISQA TEST LIVETALK & ODM & 0.154 & [-0.008, 0.064]	     & No Diff. & 0.84 & 0.81 \\
TCD VOIP            & ODM & 0.005 & [0.015, 0.068]	     & \textbf{SCOREQ} & 0.86 & 0.83 \\
TENCENT             & ODM & 0.000 & [-0.019, -0.08]      & L2 & 0.77 & 0.78 \\
TENCENT-Rev         & ODM & 0.000 & [0.079, 0.105]	     & \textbf{SCOREQ} & 0.44 & 0.35 \\

    \hline
    \end{tabularx}
  \label{tab:stats_synth}
\end{table}

\newpage
\newpage

\section*{NeurIPS Paper Checklist}

\begin{enumerate}

\item {\bf Claims}
    \item[] Question: Do the main claims made in the abstract and introduction accurately reflect the paper's contributions and scope?
    \item[] Answer: \answerYes{} 
    \item[] Justification: Our contribution is clearly stated in abstract and introduction

\item {\bf Limitations}
    \item[] Question: Does the paper discuss the limitations of the work performed by the authors?
    \item[] Answer: \answerYes{} 
    \item[] Justification: We created a separate section for limitations (Appendix \ref{appendix:limitations}). We outline the scope of our paper and the limitations of our proposed solution.

\item {\bf Theory Assumptions and Proofs}
    \item[] Question: For each theoretical result, does the paper provide the full set of assumptions and a complete (and correct) proof?
    \item[] Answer: \answerNA{} 
    \item[] Justification: Paper not including theoretical results

    \item {\bf Experimental Result Reproducibility}
    \item[] Question: Does the paper fully disclose all the information needed to reproduce the main experimental results of the paper to the extent that it affects the main claims and/or conclusions of the paper (regardless of whether the code and data are provided or not)?
    \item[] Answer: \answerYes{} 
    \item[] Justification: The paper is fully reproducible. We have provided the weights, as well as detailed instruction in Appendices \ref{appendix:architecture} and \ref{appendix:training_details} Although we will release the full code, experiments are fully reproducible with the information provided.

\item {\bf Open access to data and code}
    \item[] Question: Does the paper provide open access to the data and code, with sufficient instructions to faithfully reproduce the main experimental results, as described in supplemental material?
    \item[] Answer: \answerYes{} 
    \item[] Justification: We used both publicly available data and pre-trained models. Also, we submitted the necessary code to run the metrics. The full code will be released upon acceptance of the paper as well as a pip package. 

\item {\bf Experimental Setting/Details}
    \item[] Question: Does the paper specify all the training and test details (e.g., data splits, hyperparameters, how they were chosen, type of optimizer, etc.) necessary to understand the results?
    \item[] Answer: \answerYes{} 
    \item[] Justification: Training details are all provided in Appendix \ref{appendix:training_details} and \ref{appendix:architecture}. Other details throughout the paper. We used publicly available train/validation/test partitions.

\item {\bf Experiment Statistical Significance}
    \item[] Question: Does the paper report error bars suitably and correctly defined or other appropriate information about the statistical significance of the experiments?
    \item[] Answer: \answerYes{} 
    \item[] Justification: Bootstrapping with confidence interval are calculated to compare our metric with the standard L2 loss approach. We compute statistics on 11 test sets. Details in Appendix \ref{appendix:statistical_analysis}.

\item {\bf Experiments Compute Resources}
    \item[] Question: For each experiment, does the paper provide sufficient information on the computer resources (type of compute workers, memory, time of execution) needed to reproduce the experiments?
    \item[] Answer: \answerYes{} 
    \item[] Justification: We give details on the GPU used and number of workers in Appendix \ref{appendix:training_details}.
    
\item {\bf Code Of Ethics}
    \item[] Question: Does the research conducted in the paper conform, in every respect, with the NeurIPS Code of Ethics \url{https://neurips.cc/public/EthicsGuidelines}?
    \item[] Answer: \answerYes{} 
    \item[] Justification: Our work uses available resources that are well known by the community. No human-participants has been directly involved by us. 

\item {\bf Broader Impacts}
    \item[] Question: Does the paper discuss both potential positive societal impacts and negative societal impacts of the work performed?
    \item[] Answer: \answerYes{} 
    \item[] Justification: We created a separate section for societal impact in Appendix \ref{appendix:impact}. We believe there is no negative impact since our work is a speech quality metric. 
    
\item {\bf Safeguards}
    \item[] Question: Does the paper describe safeguards that have been put in place for responsible release of data or models that have a high risk for misuse (e.g., pretrained language models, image generators, or scraped datasets)?
    \item[] Answer: \answerNA{} 
    \item[] Justification: Our model predicts quality scores. 

\item {\bf Licenses for existing assets}
    \item[] Question: Are the creators or original owners of assets (e.g., code, data, models), used in the paper, properly credited and are the license and terms of use explicitly mentioned and properly respected?
    \item[] Answer: \answerYes{} 
    \item[] Justification: We did not provide a licence. Our work is based on publicly available assets for the research community. We cited original work. 

\item {\bf New Assets}
    \item[] Question: Are new assets introduced in the paper well documented and is the documentation provided alongside the assets?
    \item[] Answer: \answerYes{} 
    \item[] Justification: We submitted the minimal code with instructions to run 2 speech quality metrics. Upon acceptance of the paper we will release the full code and create an exhaustive README file to run experiments.

\item {\bf Crowdsourcing and Research with Human Subjects}
    \item[] Question: For crowdsourcing experiments and research with human subjects, does the paper include the full text of instructions given to participants and screenshots, if applicable, as well as details about compensation (if any)? 
    \item[] Answer: \answerNA{} 
    \item[] Justification:

\item {\bf Institutional Review Board (IRB) Approvals or Equivalent for Research with Human Subjects}
    \item[] Question: Does the paper describe potential risks incurred by study participants, whether such risks were disclosed to the subjects, and whether Institutional Review Board (IRB) approvals (or an equivalent approval/review based on the requirements of your country or institution) were obtained?
    \item[] Answer: \answerNA{} 
    \item[] Justification: 

\end{enumerate}

\end{document}